%% file: manuscript.tex
\documentclass[reprint, amsmath,amssymb,aps]{revtex4-2}

\usepackage{graphicx}
\usepackage{dcolumn}
\usepackage{bm}
\usepackage{amsmath}
\usepackage[usenames,dvipsnames]{xcolor}
\usepackage{mathtools}
\usepackage[colorlinks=true,citecolor=blue,linkcolor=blue]{hyperref}
\usepackage[caption=false]{subfig}
\usepackage{bbm}
\usepackage{comment}
\usepackage{hyperref}
\usepackage{upgreek}
\usepackage{esint}
\usepackage{soul}
\usepackage[normalem]{ ulem } 
\usepackage{floatrow}
\usepackage{verbatim}
\usepackage{braket}
\usepackage{amsmath, amssymb}
\usepackage{array}
\usepackage{algorithm}
\usepackage{algpseudocode}
\usepackage{graphicx}

\usepackage{enumitem}

\setlist[itemize]{parsep=1em, itemsep=1em} 
\setlist[enumerate]{parsep=0em, itemsep=1em} 

\newcommand{\PreserveBackslash}[1]{\let\temp=\\#1\let\\=\temp}
\newcolumntype{C}[1]{>{\PreserveBackslash\centering}p{#1}}

\newcommand{\rom}[1]{\uppercase\expandafter{\romannumeral #1\relax}}

\begin{document}

\title{Efficient compilation of quantum circuits using multi-qubit gates}

\author{Jonathan Nemirovsky}
\author{Maya Chuchem}
\author{Yotam Shapira}\email{yotam.shapira@quantum-art.tech}

\affiliation{Quantum Art, Ness Ziona 7403682, Israel}

\begin{abstract}
As quantum processors grow in scale and reliability, the need for efficient quantum gate decomposition of circuits to a set of specific available gates, becomes ever more critical. The decomposition of a particular algorithm into a sequence of these available gates is not unique. Thus, the fidelity of an algorithm's implementation can be increased by choosing an optimized decomposition. This is true both for noisy intermediate-scale quantum platforms as well as for implementation of quantum error correction schemes. Here we present a compilation scheme which implements a general-circuit decomposition to a sequence of Ising-type, long-range,  multi-qubit entangling gates, that are separated by layers of single qubit rotations. We use trapped ions as an example in which multi-qubit gates naturally arise, yet any system that has connectivity beyond nearest-neighbors may gain from our approach. We evaluate our methods using the quantum volume test over $N$ qubits. In this context, our method replaces $3N^2/2$ two-qubit gates with $2N+1$ multi-qubit gates. Furthermore, our method minimizes the magnitude of the entanglement phases, which typically enables an improved implementation fidelity, by using weaker driving fields or faster realizations. 
We numerically test our compilation and show that, compared to conventional realizations with sequential two-qubit gates, our compilations improves the logarithm of quantum volume by $20\%$ to $25\%$.
\end{abstract}

\maketitle

\section{Introduction}

Quantum computers have the potential to significantly enhance many computational processes in various technological fields.
The advent of digital quantum algorithms is contingent on a reliable performance of the underlying quantum hardware.
Indeed recent years have seen a steady improvement in accuracy and scope in implementation of many quantum algorithms \cite{shaydulin2023qaoa,willsch2024state,decross2024computational,pokharel2023demonstration,chowdhury2024enhancing}.
However, realization of canonical large-scale quantum algorithms, e.g. Shor's algorithm \cite{shor1994algorithms},
and Grover's algorithm \cite{grover1996fast},
are considered to be beyond the capabilities of contemporary quantum computers \cite{preskill2018quantum}.

An immediate approach to improve the overall performance and applicability of quantum computers is by improving the fidelity of their native gates, i.e. the set of primitive operations supported by the quantum hardware. Yet, a more comprehensive path lies in circuit compilation, i.e. the way native gates are composed in order to realize quantum circuits. Efficient compilation aims to reduce the number of low-fidelity native gates, resulting in a preferable performance of the quantum algorithm. 

Efficient compilation is important not only for noisy intermediate-scale quantum (NISQ) applications, but also for fault-tolerant quantum computation, by optimizing the realization of algorithms on the logical level, as well as optimizing the error detection and correction sequences, making the fault-tolerant threshold less stringent \cite{schwerdt2022comparing}.

The most common heuristics for efficient compilation are aimed at either minimization of two-qubit entangling gates \cite{goldfriend2024design} or minimization of non-Clifford gates such as the T-gate \cite{heyfron2018efficient,kissinger2020reducing,amy2014meet,debeaudrap2020techniques}. The former assumes that entangling gates typically have a lower fidelity and are slower compared to single qubit gates. The latter assumes that logical-qubit non-Clifford gates require the use of many physical operations \cite{knill2004fault,bravyi2005universal,rodriguez2024experimental}.

Here, we present a method for leveraging programmable multi-qubit (MQ) gates, for efficient compilation of arbitrary quantum circuits. Specifically we consider gates of the form,
\begin{equation}
	U_\text{MQ}\left(\varphi\right)=\exp\left(i\sum_{n,m=1}^{N}\varphi_{n,m}Z_n Z_m\right),\label{eqU}
\end{equation}
with $Z_n$ denoting a $Z$-Pauli operator acting on the $n$th qubit in a $N$-qubit register and $\varphi$ an arbitrary $N \times N$ real-valued coupling matrix. This MQ gate realizes a layer of commuting two-qubit $ZZ$ interactions, acting simultaneously on many, possibly overlapping, pairs of qubits. It is also equivalent to an evolution under an Ising-type Hamiltonian.

These gates utilize $\mathcal{O}\left(N^2\right)$
two-qubit entangling operations that may be realized simultaneously in a single step, potentially enabling a quadratic speed-up in runtime. Yet, it is not a-priori clear that this large amount of couplings can be used generically for improving algorithm performances, compared to the more conventional realization using sequential two-qubit gates.

Gates of the form of Eq. \eqref{eqU} are native to quantum computers based on trapped-ions
\cite{martinez2016compiling,shapira2023fast,schwerdt2024scalable,shapira2023programmable,shapira2020theory,grzesiak2020efficient,lu2023realization,wu2023qubits,wang2022fast}. Long-range entangling operations are also found in arrays of optically trapped neutral atoms \cite{evered2023high,young2021asymmetric} and atoms in optical cavities \cite{cooper2024graph}. The advantage provided by these has been shown in several specific instances, such as in compilation of an $N$-qubit Toffoli gate and Clifford circuits \cite{bravyi2022constant}, an $N$-qubit quantum Fourier transformation \cite{bassler2023synthesis} and in quantum error correction \cite{schwerdt2024scalable}.

Motivated by the fact that for fast MQ gates implemented in trapped-ions based quantum hardware, the gate duration is independent of $\varphi$ and that errors, such as photon-scattering, are not determined by the number of couplings \cite{shapira2023fast}, we apply here a new heuristic for efficient compilation, namely minimizing the number of MQ entangling layers. We derive a closed-form method that generically accomplishes this by `fusing' many pairwise interactions to single large-scale MQ gate layers. Furthermore our method reduces the drive amplitude required to realize these gates, and therefore enhances the implementation's accuracy.

Contrasting our approach with minimization of the number of two-qubit gates - we aim to generate compilations that reduce the number of MQ gates of the form of Eq. \eqref{eqU}, while imposing no restriction on the specific $\varphi_{n,m}$ phases used in each instance of it. Importantly we can make use of many simultaneous two-qubit interactions, that may even surpass the two-qubit gate count of the original circuit. For instance, a 3-qubit Toffoli gate implemented with CNOT gates requires the application of at least 6 such gates \cite{yu2013five}. Yet, in a trapped-ion quantum computer, equipped with a simultaneous all-to-all connectivity, a preferred strategy is to realize the Toffoli gate with a minimal number of MQ gates. Indeed, a Toffoli gate can be realized with just 3 MQ gates, with a total of 7 non-vanishing values of the $\varphi_{n,m}$ phases (see the SM \cite{SM}).

\begin{figure}[t]
	\includegraphics[width=\textwidth]{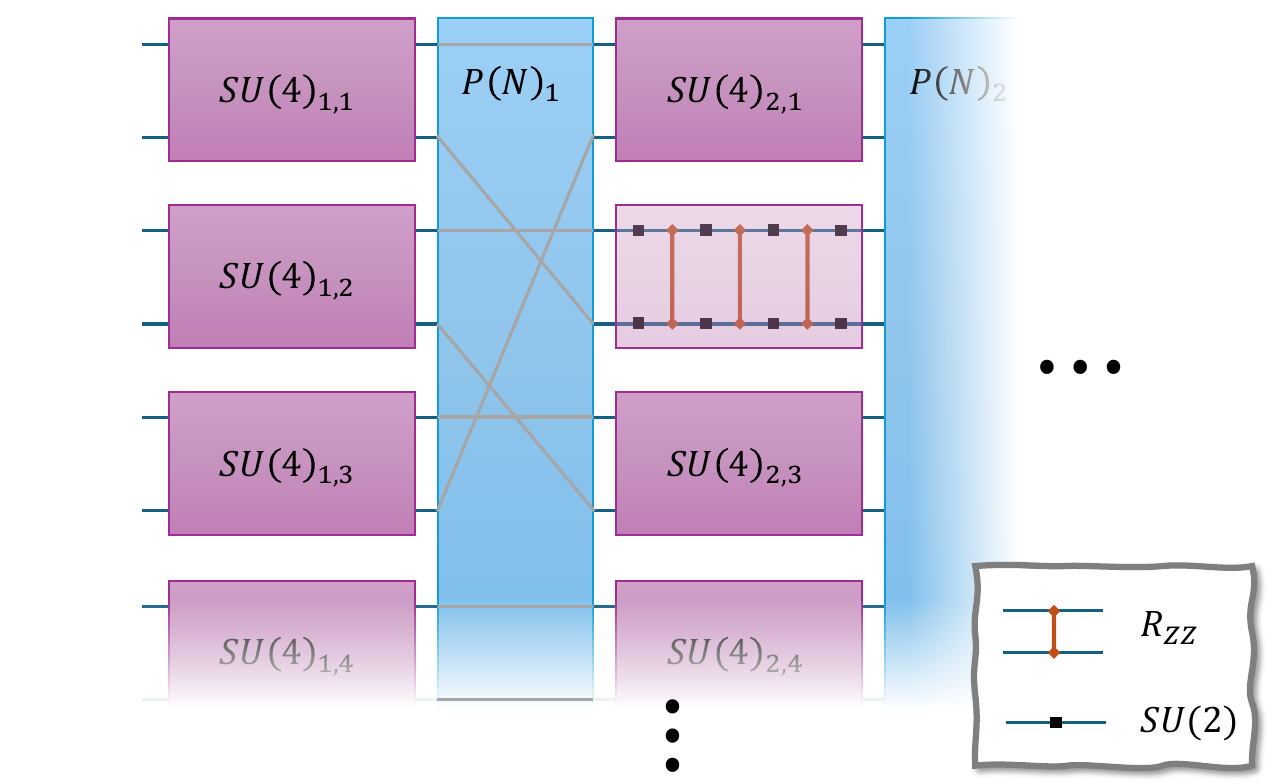}
	\caption{General structure of the quantum volume test. The circuit is constructed in layers (horizontal direction), such that in each layer the qubits are arranged in pairs, and undergo Haar random, $SU(4)$, evolution (purple). The qubits then undergo a random permutation, P(N) (blue), that randomizes the pairing for the next layer. An example for such a permutation is shown in a the first layer (gray lines). QV circuits are constructed by $N$ consecutive layers. $SU(4)$ gates are conventionally compiled (transparent) with three $R_\text{ZZ}$ gates (orange lines) and several, SU(2), single-qubit gates (black squares).}
	\label{figQV_full}
\end{figure}

To make our presentation concrete we present our methods in the context of the quantum volume (QV) test. This is a well-accepted full-system benchmark that evaluates the performance of quantum computers \cite{moll2018quantum,cross2019validating,baldwin2022reexamining,miller2022improved} and has been recently measured on various quantum hardware \cite{jurcevic2021demonstration,moses2023race,quantinuum2024quantinuum}. A QV(N) test is performed by implementing many realizations of randomized circuits that couple $N$ qubits in $N$ layers. The test is `passed' if these circuits are executed to a sufficient fidelity, quantified by a heavy-weights analysis \cite{aaronson2016complexity}.

The general structure of the randomized circuits is shown in Fig. \ref{figQV_full}. The circuit is constructed of $N$ layers, such that in each layer the $N$ qubits are divided to pairs and undergo Haar random $SU(4)$ evolution (purple, $SU\left(4\right)_{i,j}$ with $i$ the layer index and $j$ the pair index). The qubits then undergo a random permutation (cyan $P\left(N\right)_i$ with $i$ the layer index) that shuffles the qubit pairing for the next layer. 

Crucially the structure of QV circuits is universal, in the sense that any polynomial-depth quantum algorithm can as well be implemented in polynomial-depth using the QV structure. Furthermore this structure resembles some prototypical relevant quantum applications such as random circuits sampling, adiabatic optimization and variational eigensolvers \cite{cross2019validating}. Thus, our optimization for it generalizes to all circuits that are given as a set of single-qubit and two-qubit gates. 

Formally the QV of a quantum computer is defined as $\log_2 \text{QV}=\underset{N}{\operatorname{argmax}} \left[\min\left(N,d(N)\right)\right]$, where $N$ is the number of qubits in the quantum register and $d\left(N\right)$ is the circuit depth for which the test is passed. This definition loosely coincides with the complexity of classically simulating the QV circuits \cite{cross2019validating}. In the low-error limit we expect, $d\left(N\right)\simeq 1/N\epsilon_\text{eff}$, with $\epsilon_\text{eff}$ an effective error of the total operations performed per qubit per layer. By assuming that the limit on QV arises from errors then, $d(N)=N$, and we expect $\log_2 \text{QV}\simeq1/\sqrt{\epsilon}$, yielding an approximate scaling of QV with errors.

Conventional compilation of QV circuits requires the application of $3N^2/2$ two-qubit gates. However, with our method, an $N$-qubit QV circuit is compiled utilizing $2N+1$ applications of MQ gates. Furthermore, the drive power required to execute our compilation, defined precisely below, is decreased by an average of $15\%$, implying a higher realization fidelity.

In most quantum computing platforms, pairing of qubits typically requires shuttling \cite{pino2021demonstration,moses2023racetrack,bluvstein2022quantum} or swapping operations \cite{jurcevic2021demonstration}, which constitute an additional overhead on the two-qubit gate count or circuit duration. Yet, with MQ gates, pairings are trivially generated, by simply utilizing the corresponding entries of $\varphi$, that reflect the qubit-pairs. Furthermore, virtual swapping operations can be used to reduce the magnitude of entanglement phases used in the realization, as is discussed in Ref. \cite{baldwin2022reexamining}.

\section{Compilation with MQ gates}

Implementation of a general two-qubit $SU(4)$ operator, requires the application of three two-qubit gates, and arbitrary single qubit rotations, as shown in the middle operator in Fig. \ref{figQV_full} (hollowed rectangle). The specific parameters of these gates are readily computed with the well-known Cartan decomposition \cite{khaneja2001time,tucci2005introduction,blaauboer2008analytical,zhang2003geometric}, and can be given in terms of correlated $ZZ$ rotations, i.e. $R_\text{ZZ}^{\left(n,m\right)}\left(\theta\right)=\exp\left(-i\theta Z_n Z_m\right)$, such that $\theta=\pi/4$ is a fully entangling operator in its respective two-qubit subspace.

We remark that only four single qubit rotations are required for the decomposition, yet we depict 8 single qubit rotations for a simpler presentation - these excess gates can be later removed in a straightforward manner and do not impose a bottleneck on performance, thus their reduction is not considered here as an optimization goal.

\begin{figure}[t]
	\includegraphics[width=\textwidth]{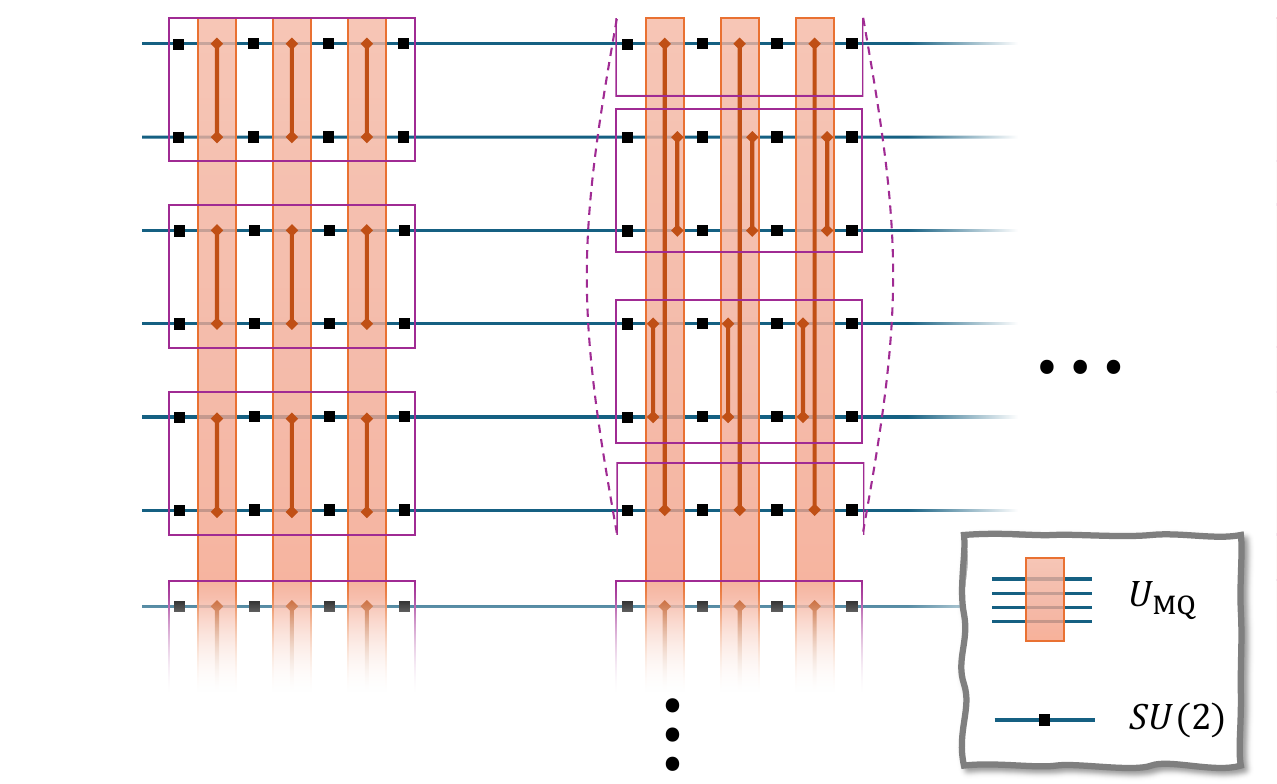}
	\caption{Straightforward compilation of a $N$-qubit quantum volume test with $3N$ uses of the MQ gate, $U_\text{MQ}$. Each MQ gate (orange shaded) realizes $ZZ$ couplings (orange lines) that reflect the qubit pairings from Fig. \ref{figQV_full}. The MQ gates naturally account for the qubit permutation layer such that no qubit shuttling or swapping is required.}
	\label{figQV_naive}
\end{figure}

A straightforward advantage of MQ gates is presented in Fig. \ref{figQV_naive}, showing a compilation of the QV circuit from Fig. \ref{figQV_full}, with $3N$ gates of the form of Eq. \eqref{eqU}. As shown, each layer is constructed of three MQ gates (orange shaded) operating on all qubits in the register and four arbitrary single-qubit rotations (black squares), per qubit. The MQ gates realize $N/2$ couplings (orange lines), that reflect the pairings of the original circuit. Furthermore, the values of the $\varphi_{n,m}$ phases of the various MQ gates, as well as the parameters of the single qubit gates, are deduced from the same Cartan decomposition used for each $SU(4)$. Crucially, with the use of MQ gates the permutation layers shown in Fig. \ref{figQV_full} are no longer required and are simply reflected by the choice of $\varphi_{n,m}$ phases of succeeding layers.

\begin{figure}[t]
	\includegraphics[width=\textwidth]{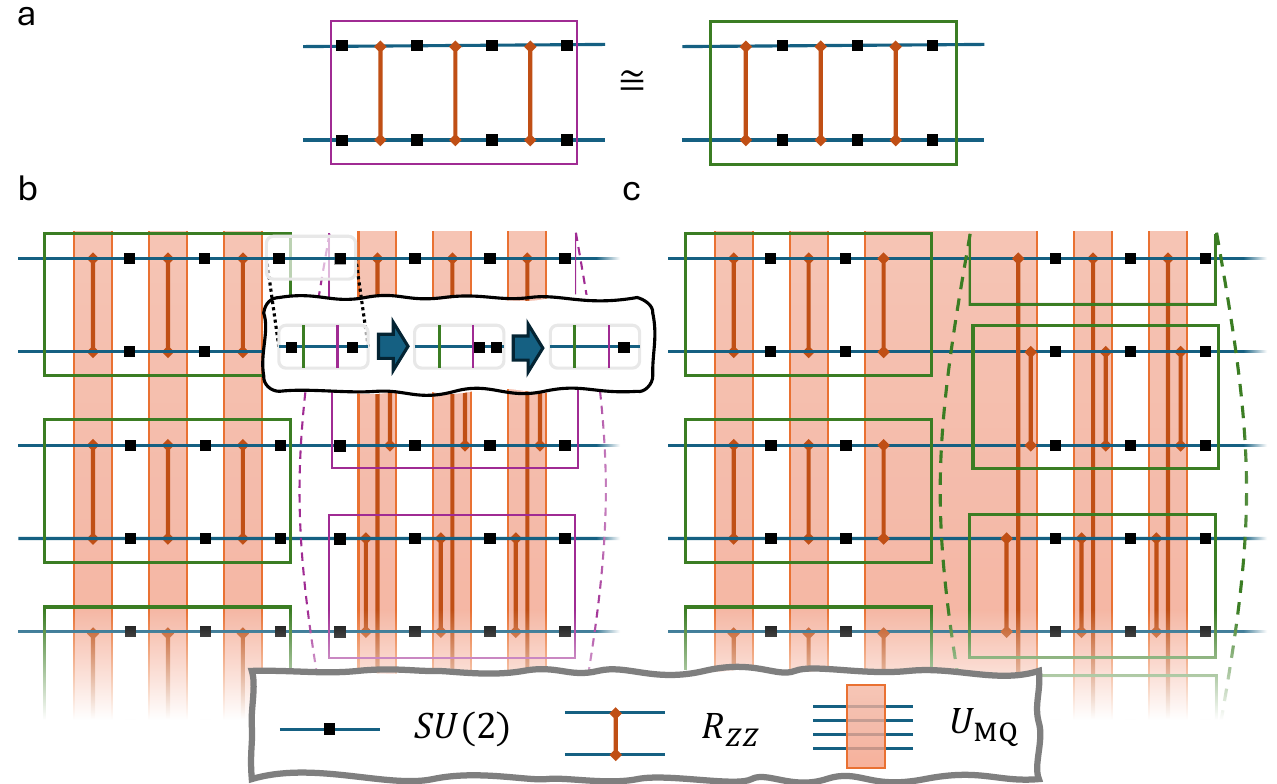}
	\caption{Efficient compilation of the quantum volume test. a. The conventional Cartan decomposition (left, in purple box) can be converted to a left-handed decomposition (right, in green box) in which the first gate in the decomposition is an $R_\text{ZZ}$ gate (orange line). b. To obtain an efficient compilation we first LH decompose the first layer of $SU\left(4\right)$ blocks (green blocks) and `push' the last single qubit rotations in each block to the next layer (inset). c. The next layer is LH decomposed such that the interface between every two layers is made of $R_\text{ZZ}$ gates, which can be `fused' to a single MQ gate. This ultimately reduces the MQ gate count from $3N$ to $2N+1$.}
	\label{figLemma}
\end{figure}

Our compilation is further optimized by considering the lemma shown in Fig. \ref{figLemma}a. Specifically, we perform a `left-handed' (LH) variant of the Cartan decomposition, in which the first operation of every $SU\left(4\right)$ block is a $R_\text{ZZ}$ gate, with no more than three $R_\text{ZZ}$ gates used per $SU\left(4\right)$ block. The method to obtain LH-decompositions is detailed below.  

Indeed, the LH-decomposition enables further minimization of the MQ gate count by the following procedure: The first (vertical) layer in the circuit is LH decomposed, shown in Fig. \ref{figQV_full}b (green blocks). With this, each $SU\left(4\right)$ has single-qubit rotations at its (rightmost) end (black squares). These are `pushed' and absorbed by the $SU\left(4\right)$ operation of the next layer (inset), such that each block now ends with a $R_\text{ZZ}$ gate. The next layer is then LH decomposed (green blocks), and its first $R_\text{ZZ}$ gate can be fused with the last $R_\text{ZZ}$ gate of the previous layer to a MQ gate (orange shading), shown in Fig. \ref{figQV_full}c. Again, the last single qubit gates of these blocks are pushed and merged to the next layer. This process repeats for the whole circuit, so that each layer (besides the last) is replaced with two MQ gates and two layers of single qubit gates. To realize the last layer of the QV circuit an additional MQ gate layer is required. Hence, the compilation of the QV circuit is realized with $2N+1$ MQ gates, instead of $3N$. 

We detail our method for obtaining LH decompositions, as in Fig. \ref{figLemma}a. For this we recall the Cartan decomposition \cite{tucci2005introduction,blaauboer2008analytical}, in which a target two-qubit unitary $U\in SU\left(4\right)$, is rotated to the so-called `magic' basis (MB) by the unitary matrix $M$, in which it abides the decomposition $M^\dagger U M=O_2^T D O_1$ with $O_n$ orthonormal real-valued matrices and $D=\exp\left[i\left(\theta_0 1\otimes1+\theta_{xx} Z\otimes1 + \theta_{yy} 1\otimes Z +\theta_{zz} Z\otimes Z\right)\right]$, a diagonal unitary matrix.

The Cartan decomposition operates by rotating $U$ back to its original basis using, $MO_nM^\dagger=S_{n,1}\otimes S_{n,2}$ such that $S_{n,j}\in SU\left(2\right)$ are single-qubit rotations and $T\equiv M D M^\dagger=\exp\left[i\left(\theta_0 1\otimes1+\theta_{xx} X\otimes X + \theta_{yy} Y\otimes Y +\theta_{zz} Z\otimes Z\right)\right]$, resulting with the decomposition $U=\text{Cartan}\left(U\right)=\left(S_{2,1}\otimes S_{2,2}\right)T\left(S_{1,1}\otimes S_{1,2}\right)$, i.e. up to a global phase a leading layer of single-qubit gates, then $R_{XX}$, $R_{YY}$ and $R_{ZZ}$ gates, and a trailing layer of additional single qubit gates. clearly $R_{XX}$ and $R_{YY}$ can both be rotated to $R_{ZZ}$ gates by using additional, pre-determined, single-qubit gates before and after these gates are applied, resulting in the structure highlighted in Fig. \ref{figLemma}a.

\label{sec:LH_decomp}
To perform our LH decomposition we define the Cartan volume (CV), as $\text{CV}\left(U\right)=\sin\left(2\theta_{xx}\right)\sin\left(2\theta_{yy}\right)\sin\left(2\theta_{zz}\right)$, with the entanglement phases obtained by the Cartan decomposition of $U$ (the notion of volume is apparent at small angles). Equivalently, CV is an invariant that takes the simple algebraic form, 
\begin{equation}
	\text{CV}\left(U\right)=-\frac{1}{4}\text{Im}\left( \text{Tr} \left( U_\text{MB}^TU_\text{MB} \right) \right),\label{eqCarAlg}
\end{equation}
with $U_\text{MB}=M^\dagger U M$, i.e., $U$ rotated to the MB (see proof in the SM \cite{SM}).

We consider the function $v\left(\xi \right)\equiv\text{CV}\left(U e^{-i \xi ZZ }\right)$, which due to Eq. \eqref{eqCarAlg} yields,
$v\left(\xi \right)=a(U)\text{sin}(2(\xi-\xi_0(U))$ with $a(U)$ and $\xi_0(U)$ being simple algebraic expressions of the entries of $U$, that are easy to compute \cite{SM}.
Next, considering the zero, $\xi_0$, of $v\left(\xi \right)$, it is apparent that $\text{Cartan}\left(U e^{-i \xi_0 ZZ } \right)$ has 
only two non-trivial non-zero entanglement phases. Thus, the LH decomposition of $U$ is given by,
\begin{equation}
	U=\text{LH}\left(U\right)=\text{Cartan}\left(U e^{-i \xi_0 Z\otimes Z } \right)e^{i \xi_0 Z\otimes Z }.\label{eqLeftHanded}
\end{equation}
We remark that in a similar fashion we can form an analogous right-handed decomposition.
Furthermore, for $|a(U)| > 0$ there is precisely one  $\xi_0$ solution in the interval $\left(-\pi/4,\pi/4\right]$. 

\section{Entanglement phases and drive amplitudes optimization}

Typically small phase values are associated with a higher implementation fidelity of the quantum hardware, as they correspond to faster realizations or small driving fields. Specifically in context of MQ gates for Raman driven qubits encoded in trapped ions, the nuclear-norm 
$\text{nuc}\left(\varphi\right) \equiv  \sum_k |\lambda_k|$, where $\lambda_k$s are eigenvalues of the $\varphi$ matrix, with all its elements taken in absolute value, is associated with the laser power required to drive the gate. In turn, laser power corresponds to errors due to spontaneous photon scattering and depolarization of the qubits \cite{shapira2023fast}. Thus, we are interested not only in reducing the MQ gate count, but also the total nuclear-norm sum over all MQ gates.

We perform this additional optimization by `injecting' identity operators in the form of single qubit rotations and their inverse, between every two layers, for all qubits. The rotations are then `pushed' back and forward in order to reduce the nuclear norm of the MQ gates that realize these layers (see the SM \cite{SM}).

Importantly, the entanglement phases of the LH decomposition and the subsequent nuclear norm of the MQ gates it underlies can be significantly impacted by this optimization procedure. This feature is, at first, surprising as the Cartan decomposition is trivially unaffected by prior or post rotations, as these merely change its single-qubit $SU(2)$ operators, while its entanglement phases remain unaffected.

Moreover, it turns out that optimization using $R_\text{Y}$ gates at the beginning of each $SU(4)$ block, suffices to realize the minimal magnitude of entanglement phases, $\left|\theta_{xx}\right|+\left|\theta_{yy}\right|+\left|\theta_{zz}\right|$, where $\theta_{nn}$ are the entanglement phases of the corresponding Cartan decomposition. This leads to a significant advantage of utilizing the LH decomposition scheme in compilation of MQ gates, as the nuclear norm of the combined optimized MQ gates (obtained by the LH compilation procedure) is typically significantly smaller than the total sum of disjoint Cartan decompositions.

While global optimization of the total nuclear norm of small QV circuits is manageable at a reasonable computation time, the overall complexity of large QV circuit's optimization is impractical with current computation means. Moreover, the QV test `prohibits' utilizing classical computational resources that scale exponentially with $N$. It is therefore essential to optimize the compilation of large circuits with efficient algorithms of polynomial complexity.

In the following we present such an optimization scheme, based on local layer-by-layer optimization, that scales with polynomial complexity. While our scheme leads to a local minima, rather than a global one, it still yields a significant reduction, typically of $\sim15\%$ in the total nuclear norm. For example, we sampled $r=100$ random QV circuit for $N=30$ and observe that the ratio between the total nuclear norm over all MQ gate between the LH-optimized compilation and the Cartan compilation yields on average $0.85$ with a standard deviation of $0.004$. 

Importantly, the average compilation time of large circuits is manageable - e.g., the optimization run-time on an ordinary laptop PC were: 3.2, 7.1 and 26.5 minutes for QV circuits with 30, 40 and 60 qubits, respectively.

The optimization is performed as follows. We iterate over the layers of the circuit from layer $\ell=1$ to $N-1$. At each iteration we assume layer $\ell$ is already LH decomposed and terminates with $R_{ZZ}$ gates and layer $\ell+1$ is not. We optimize over $N$ single qubit $R_Y$ rotations between the layers, such that rotations are `pushed forward' to the blocks at layer $\ell+1$ and inverse rotations are `pulled back' into the blocks of layer $\ell$. With this, layer $\ell$ is fixed at each iteration. After the iterations conclude we merge all relevant $R_{ZZ}$ gates and the compilation is complete. We remark that `pulling' $R_{Y}$ rotations into LH decomposed blocks is non-trivial, a closed form method to do so is given in the SM \cite{SM}.

Additional optimization heuristics may be employed, such as adding two-qubit identities in the form of consecutive SWAP gates at the end of an $SU(4)$ block. This is useful since in practice one of the gates is implemented using entanglement gates and absorbed in the LH decomposition, while the other is implemented `virtually' by relabeling qubit indices in subsequent gates. Such a method has also been adopted in Ref. \cite{baldwin2022reexamining}.

\section{Noise models}

To test the expected performance of our compilation, we consider relevant noise models. Namely, qubit depolarization due to photon scattering and qubit phase noise, both assumed to increase linearly with the gate duration. These models are chosen due to their relevance to high-fidelity entanglement gates in trapped-ions qubits \cite{ballance2016high}. Here, we focus solely on errors in entanglement operations as single-qubit gate errors are generally subdominant \cite{smith2024single}.

Specifically, for qubits encoded in the atomic state of a trapped ion and coupled by a Raman transition, depolarization is generated by off-resonance photon scattering from short-lived states, which occur at a rate that is linear in the driving field's power. For MQ gates the depolarization probability has been shown \cite{shapira2023fast} to be proportional to the nuclear norm of coupling matrix $\varphi$ of Eq. \eqref{eqU}, i.e. it is given by the $L_1$ norm of the eigenvalues of $\varphi$ with its elements taken in absolute value. Since photon scattering is a qubit-local effect the local drive power per qubit must be considered. Here it is taken into account using the relative qubit participation $\alpha_m$, defined as,
\begin{equation}
	\alpha_n = \sum_m|{\varphi_{n,m}}|/\sum_{m,s}|{\varphi_{m,s}}|.\label{eqParticipation}
\end{equation}
The MQ error is `gauged' by a two-qubit error parameter, $p^{\text{TQ}}_{\text{depol}}$, which is an external parameter determining the error-probability per qubit of a fictitious fully entangling two-qubit gate, i.e. its entanglement phase is $\varphi_2=\pi/4$. We model Raman scattering by the formula for off-resonance excitation,
\begin{equation}
	p^{\text{TQ}}_{\text{depol}}=\frac{\frac{1}{2}{\text{nuc}}(\varphi_2)}{\frac{1}{2}{\text{nuc}}(\varphi_2)+\Delta^2},
\end{equation}
with the factors of 1/2 coming from the equal participation of the two ions and $\Delta$ marking an effective detuning from the short-lived level.
Inverting this relation, we obtain, $\Delta^2=\text{nuc}(\varphi_2)(1-p^{\text{TQ}}_{\text{depol}})/(2p^{\text{TQ}}_{\text{depol}})$. With this, $\Delta^2$ is used to convert the error rate between two-qubit and MQ gates. Specifically, the probability of depolarization of the $n$th qubit in a MQ gate is given by,
\begin{equation}
 p^{\text{MQ}}_{\text{depol},n} =  \frac{\alpha_n{\text{nuc}}(\varphi)}{\alpha_n{\text{nuc}}(\varphi)+\Delta^2} =\frac{4 {\text{nuc}}(\varphi) p^{\text{TQ}}_{\text{depol}} \alpha_n}{1+p^{\text{TQ}}_{\text{depol}}(4{\text{nuc}}(\varphi)\alpha_n-1)}. \label{eqPn}
\end{equation}

Clearly, this mechanism holds as well for two-qubit gates with general phases. With this definition the error rates, $p^{\text{MQ}}_{\text{depol},n}\left(\varphi\right)$, of a MQ gate that couples independent pairs of qubits, i.e. each row of $\varphi$ has at most only one non-zero value (e.g., the left-most MQ gate in Fig. \ref{figLemma}c) is identical to that of an equivalent sequential two-qubit gate implementation. However such an equality will not hold for general $\varphi$s where qubits are coupled to more than one qubit, in which case the MQ error will be lower.

For the assumed square-root relation between errors and $\log_2\text{QV}$, above, we expect our compilation to increase $\log_2\text{QV}$ by $1/\sqrt{0.85}-1\approx8.5\%$, compared to a sequential two-qubit gate implementation, under depolarization noise.

We now turn to the second noise model, dephasing, which is relatively simpler. Here we assume a `background' dephasing channel due to, e.g. magnetic field noise, which only depends on gate duration. Thus, in this model each qubit that takes part in an entanglement gate undergoes local dephasing with probability $p^{\text{TQ}}_\text{dephase}$. Clearly such a model puts an emphasis on the number of entanglement operations per ion. Similarly, for the assumed square-root relation between errors and $\log_2\text{QV}$, above, we expect our compilation to increase $\log_2\text{QV}$ by $1/\sqrt{2/3}-1\approx23\%$, compared to a sequential two-qubit gate implementation, under dephasing noise.

\section{Numerical benchmarks}

We benchmark our compilation by numerically performing the QV test and comparing its results to a conventional implementation with sequential two-qubit gates. We consider the noise models described above - depolarization and dephasing. In each of these we scan different noise probabilities and observe the resulting $\log_2\text{QV}$ of each compilation method.

Numerical simulations are performed by sampling $n_c \geq 100$ QV circuits for various values of $N$. Each QV circuit is compiled with our compilation method, using MQ gates or with two-qubit gates. 

Noise is implemented by calculating the error probabilities of each entanglement gate, according to the noise model, and sampling and injecting corresponding Pauli noise channels, prior to the entanglement gates. For example, to simulate depolarization noise at the $n$th qubit, random Pauli operators acting on the $n$th qubit are added  prior to each entanglement gate acting on this qubit, with probability $p^{\text{MQ}}_{\text{depol},n}(\varphi)$ in Eq. \eqref{eqPn}. 

Finally, our simulations are performed with a quantum circuit state simulator on an NVIDIA H100 GPU, running CUDA-Q \cite{cuda-q}.

In order to estimate the passing qubit number, $N$, i.e. $\log_2\text{QV}$, as function of the effective two-qubit gate error, $p^\text{TQ}$, we sampled the error values using a binary search method, up to a resolution, $\Delta p = 10^{-3}\min\left\{(10/N)^2,1\right\}$, i.e. it is set to $10^{-3}$ for $N\leq10$ and decreases quadratically with $N$. Similarly the number of total shots is $r=10^4\max\left\{(N/10)^4 ,1\right\}$, i.e. it is set to $10^4$ for $N\leq10$ and increases quartically with $N$.

\begin{figure}[t]
	\includegraphics[width=\textwidth]{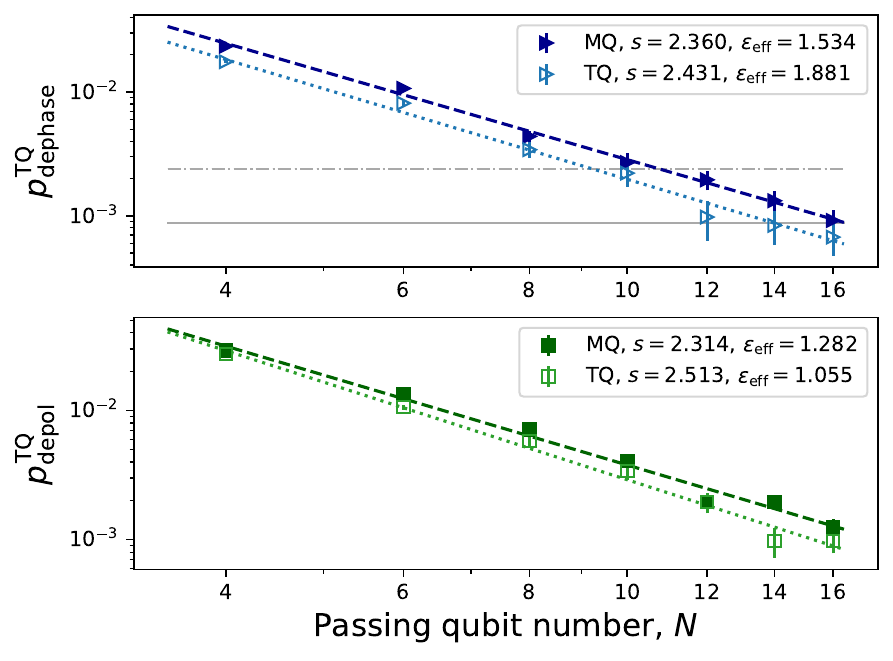}
	\caption{A comparison of QV passing qubit number, $N$, i.e. $\log_2\text{QV}$ (horizontal, log-scale) and effective error probability (vertical, log-scale), under a dephasing noise model (top) and depolarization (bottom), for our MQ compilation method (dark, full markers) and a realization with sequential two-qubit gates (light, hollow markers). For both models, our compilation provides a preferable performance and a higher QV. The numerical results (markers) are fitted to an analytical expectation, $p^\text{TQ}=1/\left(\epsilon_\text{eff} N^s\right)$ (dashed), with the fitting parameters shown in the legend and confidence intervals detailed in the main text. As an example we consider dephasing errors of $p^\text{TQ}_\text{dephase}=2.4\cdot10^{-3}$ (top, gray dot-dashed) and $p^\text{TQ}_\text{dephase}=8.7\cdot10^{-4}$ for which the passing qubit number due to our compilation is larger by 2 and 4, respectively, compared to an implementation with two-qubit gates.}
	\label{figResults}
\end{figure}

We present the results of these numerical evaluations in Fig. \ref{figResults}, with passing qubit number, $N$, i.e. $\log_2\text{QV}$, in the horizontal axis (log scale) and the effective two-qubit gate error in the  vertical axis (log scale), with $p^{\text{TQ}}_{\text{dephase}}$ for the dephasing noise model (top, blue, triangles) and $p^{\text{TQ}}_{\text{depol}}$ for the depolarization noise model (bottom, green, squares). The numerical benchmarks show an advantage for our compilation using MQ gates (full markers) over a realization with two-qubit gates (hollow markers), throughout the tested regimes and for both noise models. 

As an example, we consider a dephasing noise probability, $p^\text{TQ}_\text{dephase}=2.4\cdot10^{-3}$ (top, gray dot-dashed), for which $\log_2\left(\text{QV}\right)=8$ using a two-qubit gates implementation, while with our compilation we have $\log_2\left(\text{QV}\right)=10$. Similarly, for a lower noise level of $p^\text{TQ}_\text{dephase}=8.7\cdot10^{-4}$ (top, gray line), $\log_2\left(\text{QV}\right)=12$ using a two-qubit gates implementation, while with our compilation we have $\log_2\left(\text{QV}\right)=16$, indicating that the advantage of our method grows as the errors decrease.

The numerical results are accompanied by a fit to a power law, $p^\text{TQ}=1/\left(\epsilon_\text{eff} N^s\right)$. We expect an approximate quadratic scaling, i.e. $s\approx2$. The fitting results are shown graphically in Fig. \ref{figResults} (dashed) and detailed in Table \ref{tab:Fits}. The coefficient of determination for all of the fits is $R^2>0.98$. In addition, the Table shows the expected $\log_2\left(\text{QV}\right)$ due to a two-qubit error, $p^\text{TQ}=10^{-4}$, which is feasible in contemporary systems. At this error level, our compilation with MQ gates provides an advantage over a realization in two-qubit gates of $\sim21\%$ and $\sim25\%$ for dephasing and depolarization noise, respectively. As quantum computers are limited by their errors, these factors directly impact the overall system performance.

\begin{figure}[t]
	\includegraphics[width=\textwidth]{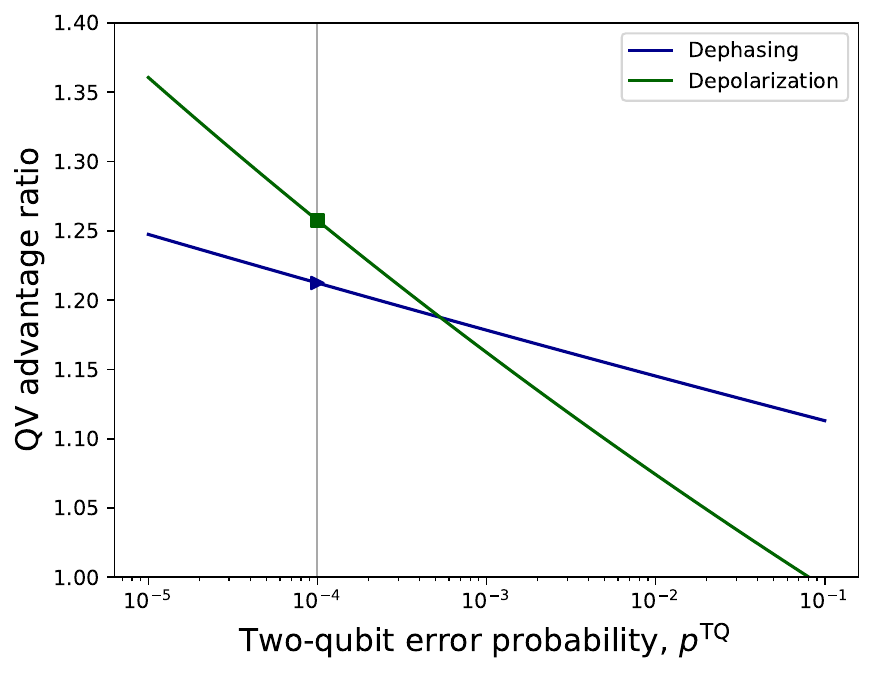}
	\caption{QV advantage ratio, $\log_2\left(\text{QV}^\text{MQ}\right)/\log_2\left(\text{QV}^\text{TQ}\right)$, extrapolated from our fits. Under depolarization noise (green) our compilation method shows an advantage regardless of the error rate. For dephasing noise (blue) our compilation method shows an advantage, below a threshold error rate, $p^\text{TQ}_\text{depol,th}\approx8.1\cdot10^{-2}$. For both models the advantage increases as errors decrease.}
	\label{figResults2}
\end{figure}

The advantage gained by our compilation is further emphasized in Fig. \ref{figResults2}, showing the QV advantage ratio, defined as $\log_2\left(\text{QV}^\text{MQ}\right)/\log_2\left(\text{QV}^\text{TQ}\right)$  (vertical), as a function of the two-qubit error probability, $p^\text{TQ}$ (horizontal, log-scale), according to our fits above. The results for $p^\text{TQ}=10^{-4}$, discussed above, are highlighted (vertical line and points). For both noise models the fitted values of $s$ for our MQ method are smaller that with an implementation with two-qubit gates. This underlies the increase in the QV advantage ratio as errors decrease. For dephasing noise the fitted value of $\epsilon_\text{eff}$ is also smaller, meaning that an advantage is maintained regardless of noise. For depolarization noise,  $\epsilon_\text{eff}$ is larger with our method, meaning that under this noise model our compilation is advantageous below a threshold error rate, which evaluates to a rather high value of $p^\text{TQ}_\text{depol,th}\approx8.1\cdot10^{-2}$ (at this value $\log_2\left(\text{QV}\right)=2$).

\begin{table}[h!]
	\centering
	\begin{tabular}{|c|c|c|c|}
		\hline
		\textbf{Model} & $s$ & $\epsilon_\text{eff}$ & $\log_2\left(\text{QV}\right)$ \\ \hline
		Dephasing TQ        & $2.431(1)$  & $1.881(3)$ &  34     \\ \hline
		Dephasing MQ        & $2.360(1)$  & $1.534(2)$ &  41     \\ \hline
		Depolarization TQ   & $2.513(1)$  & $1.055(2)$ &  38     \\ \hline
		Depolarization MQ   & $2.314(1)$  & $1.282(2)$ &  48     \\ \hline
	\end{tabular}
	\caption{Fit results for the data shown in Fig. \ref{figResults}, according to the model $p^\text{TQ}=1/\left(\epsilon_\text{eff} N^s\right)$. Confidence intervals are shown in parenthesis as the least-significant digit of the fitted values. The expected quadratic scaling is approximately met in all models, $s\approx2$. For both noise models our compilation scales better than an implementation with two-qubit gates, i.e., $s$ is smaller, yielding an advantage that increase with decreasing errors (see Fig. \ref{figResults2}). As an example, we consider the $\log_2\left(\text{QV}\right)$ attained in each compilation and noise models at  $p^\text{TQ}=10^{-4}$ (right-most column) showing an increase in the passing qubit number of 7 (10) for qubit deaphsing (depolarization).}
	\label{tab:Fits}
\end{table}

\section{Conclusions}

In conclusion, we presented a compilation method that utilizes MQ gates in order to significantly reduce the number of entanglement gates required to realize QV tests, from $3N^2/2$ to $2N+1$. Furthermore, our method optimizes the total nuclear norm, a metric of the magnitude of entanglement phases, which corresponds to a higher implementation fidelity of the circuits. 

We have tested our compilation numerically, and contrasted it with a conventional implementation with sequential two-qubit gates, under various noise models. We have demonstrated an increase of $20\%$ to $25\%$ in $\log_2\text{QV}$ for hardware that supports MQ gates and operates at an effective two-qubit gate infidelity of $10^{-4}$. This increase implies an improved performance gained from the quantum hardware.

Lastly, we remark that the improvements given by our compilation still do not utilize the full potential of MQ gates, since in each layer only $N$ out of the $N\left(N-1\right)/2$ entanglement phases are currently populated, motivating further research of the utilization of MQ gates.

\begin{acknowledgments}
	We thank Roee Ozeri and Amit Ben Kish for valuable conversations and insights. We thank Aviv Farhi and Noa Pariente for their support in providing the HPC infrastructure. 
\end{acknowledgments}

\bibliographystyle{unsrt}
\bibliography{references}

\cleardoublepage
\onecolumngrid
{
	\begin{center}
		{\large \bfseries Supplemental material \par}
	\end{center}	
 	\bigskip
 	\setcounter{section}{0}
 	{\input{supplementary.tex}}
	\unskip
}

\end{document}

%% file: supplementary.tex
\section{Efficient realization of Tofolli gate}
We contrast our optimization approach, that reduces the total number of MQ layers, with alternative compilations that aim to reduce the number of two-qubit gates. By way of example we consider the Toffoli gate. It is known that five two-qubit gate are necessary for implementing the Toffoli gate \cite{yu2013five} (or six CNOT gates). However, a compilation of a 3-qubit Toffoli gate with MQ gates is given by,
\begin{equation}
	\text{Tof}_3=e^{-i\frac{\pi}{8}\left(Z_1 +Z_2 +Z_3\right)}\underline{e^{i\frac{\pi}{8}\left(Z_1 Z_2 +Z_1 Z_3 + Z_2 Z_3\right)}} H_1 e^{-i\frac{\pi}{4}\left(Z_2+Z_3\right)}\underline{e^{i\frac{\pi}{4}\left(Z_1 Z_2+ Z_1 Z_3\right)}} H_1 e^{i\frac{\pi}{8}Z_1} H_1 e^{-i\frac{\pi}{4}\left(Z_2+Z_3\right)}\underline{e^{i\frac{\pi}{4}\left(Z_1 Z_2 +Z_1 Z_3\right)}},\label{eqSupTof}
\end{equation}
with $H_1$ a Hadamard gate on the first qubit. Here only three layers of MQ gates are used (underlined). Note that the couplings used to generate the gate do not correspond to five two-qubit gates. In fact, in total, 7 different two-qubit couplings are employed. That is, the number of entanglement layers is reduced at the expense of more two-qubit couplings. A similar case for Toffoli gates over larger qubit registers is found in Ref. \cite{bravyi2022constant}.

\section{Existence of LH decomposition and properties of the Cartan volume}
For the sake of clarity, we provide two proofs for the existence of LH decomposition. The first proof, in section \ref{sec:topological_proof} provides insights 
and clarifies the topological property that guaranty the existence of LH decomposition. In section \ref{sec:algebraic_proof} we provide an additional, by-construction, proof which is based on an algebraic expression for Cartan volume $\text{CV}(Ue^{-i\xi ZZ})$. This proof, while somewhat technical, offers a quick and straightforward solution method along with an algebraic characterization of all the solutions.
\subsection{Proof of existence based on topological properties of Cartan decomposition}
\label{sec:topological_proof}
We prove the existence of the LH decomposition, i.e. that any $U\in SU\left(4\right)$ has a decomposition into a leading $R_\text{ZZ}$ gate followed by arbitrary single-qubit rotations and, at most, two additional $R_\text{ZZ}$ gates. Following the formalism of \cite{tucci2005introduction,blaauboer2008analytical}, we recall that the Cartan decomposition yields up to a global phase the form,
\begin{equation}
	U=e^{i \left(A_0+A_1\right)}e^{i\left(\theta_{xx}XX+\theta_{yy}YY+\theta_{zz}ZZ\right)}e^{i \left(B_0+B_1\right)},\label{eqSupCartan}
\end{equation}
with $A_n$ and $B_n$ single-qubit hermitian operators acting on the $n$th qubit and
$ 0 \le \theta_{xx},\theta_{yy},\theta_{zz} < \pi/2 $.

We define a homotopy $W:[0,\pi/2]\times[0,1]\rightarrow SU(4)$,
\begin{equation}
	W\left(\xi,h\right) \equiv Ue^{i\left(h-1\right)\left(B_0+B_1\right)}e^{-i\xi ZZ},\label{eqSupW}
\end{equation}
such that $W\left(0,0\right)=U$.
Clearly this transformation is continuous as it is only a multiplication of $4\times4$ matrices
that are continuous in $h$ and $\xi$.

We consider the transformation to the `magic' basis \cite{tucci2005introduction,blaauboer2008analytical},
$W_\text{MB}\left(\xi,h\right)\equiv M^\dagger	W\left(\xi,h\right)M$, with $M$ being the `magic' unitary basis transformation,
\begin{equation}
	M \equiv
	\frac{1}{\sqrt{2}}\begin{bmatrix}
		1 & 0 & 0 & i \\
		0 & i & 1 & 0 \\
		0 & i & -1 & 0 \\
		1 & 0 & 0 & -i
	\end{bmatrix}.
\end{equation}
Yielding (see \cite{tucci2005introduction,blaauboer2008analytical}),
\begin{equation}
	W_\text{MB}\left(\xi,h \right)=O_1^T\left(\xi,h \right)D\left(\xi,h\right)O_2\left(\xi,h\right)=
	O_1^T\left(0,h\right)D\left(0,h\right)O_2\left(0,h\right)e^{-i\xi ZZ},\label{eqSubWM}
\end{equation}
with $D\left(\xi,h\right)$ being a diagonal $4\times 4$ matrix and $O\left(\xi,h \right)_{1,2}$
being orthogonal $4\times 4$ matrices.
Note that in equation \eqref{eqSubWM} we have used the commutation relation $\left[M,ZZ\right]=0$.

We then consider the product $W_\text{MB}\left(\xi,h\right)^T W_\text{MB}\left(\xi,h\right)$.
On one hand we obtain,
\begin{equation}
	W_\text{MB}\left(\xi,h\right)^T W_\text{MB}\left(\xi,h\right)=O_2^T\left(\xi,h\right)D^2\left(\xi,h\right)O_2
	\left(\xi,h\right),\label{eqSupOneHand}
\end{equation}
and on the other hand we obtain,
\begin{equation}
	W_\text{MB}\left(\xi,h\right)^T W_\text{MB}\left(\xi,h\right)=e^{-i\xi ZZ}O_2^T\left(0,h\right)D^2\left(0,h\right)O_2
	\left(0,h\right)e^{-i\xi ZZ}. \label{eqSupOtherHand}
\end{equation}
The former, Eq. \eqref{eqSupOneHand}, shows that the spectrum of $W_\text{MB}\left(h,\xi\right)^T W_\text{MB}\left(h,\xi\right)$
is the spectrum of $D^2\left(\xi,h\right)$,
while the latter, Eq. \eqref{eqSupOtherHand} shows that this spectrum is continuous in $\left(\xi,h\right)$,
as it is given by a multiplication of $4\times4$ matrices that are smooth in $\xi$ and $h$.

The entanglement phases are given by \cite{tucci2005introduction,blaauboer2008analytical},
\begin{equation}
	\begin{split}
		\theta_{xx}\left(\xi,h\right)&=-\frac{i}{4}\log\left(\lambda_1^{2}\lambda_2^{2}\lambda_3^{-2}\lambda_4^{-2}\right)+\frac{\pi}{2}m_{xx}\\
		\theta_{yy}\left(\xi,h\right)&=-\frac{i}{4}\log\left(\lambda_1^{-2}\lambda_2^{2}\lambda_3^{-2}\lambda_4^{2}\right)+\frac{\pi}{2}m_{yy}\\
		\theta_{zz}\left(\xi,h\right)&=-\frac{i}{4}\log\left(\lambda_1^{2}\lambda_2^{-2}\lambda_3^{-2}\lambda_4^{2}\right)+\frac{\pi}{2}m_{zz},\\
	\end{split}
\end{equation}
where $\lambda_n=\lambda_n\left(\xi,h\right)$ the eigenvalues of the diaghonal matrix $D^2\left(\xi,h\right)$ and
$m_{xx}, m_{yy}, m_{zz}\in\mathbb{Z}$ being integer numbers.
The entanglement phases $\theta_{xx}, \theta_{yy}$ and $\theta_{zz}$ can be chosen to be continuous as
the $\lambda_n$s are of the form $\exp\left(i\phi_n\right)$ and recall that the complex-valued logarithm has
a branch cut at $\pm\pi$ which is made continuous with a proper choice of the integer values
$m_{xx}, m_{yy}, m_{zz}\in\mathbb{Z}$.
These integer phase factors may be added to the entanglement phases since they constitute local rotation, e.g.
$R_\text{ZZ}\left(\pi/2\right)=ZZ$, such that choosing $m_{zz}$ to make $\theta_{zz}$ continuous will correspondingly change the $A_n$s and $B_n$s.

Next we notice that for $h=0$ the problem is trivial as a
$\theta_{zz}\left(\xi,h=0\right)=\theta_{zz}\left(\xi=0,h=0\right)-\xi$.
Furthermore, at an arbitrary value of $0\le h \le 1 $ and $\xi=\frac{\pi}{2}$ we have,
\begin{equation}
	W\left(\xi=\frac{\pi}{2},h\right)=U e^{i\left(h-1\right)\left(B_0+B_1\right)}ZZ.
\end{equation}
Utilizing this and noticing that the term $e^{i\left(h-1\right)\left(B_0+B_1\right)}ZZ$
in the right hand-side is entirely composed of single-qubit operators
we conclude that,
\begin{equation}
	\theta_{zz}\left(\frac{\pi}{2},h\right)-\theta_{zz}\left(\xi=0,h\right)=k\left(h\right)\frac{\pi}{2},
\end{equation}
with $k\left(h\right)\in\mathbb{Z}$.
Moreover, continuity of $\theta_{zz}$ as a function of $h$ guarantees that $k$ is also continuous,
and since it is a function with integer values, it must be constant for all $h$. Hence we conclude that,
$k\left(h\right)=k\left(h=0\right)=-1$.
Lastly, Since $\theta_{zz}\left(h,\xi\right)$ is continuous in $\xi$,
then according to the intermediate value theorem,
there must be at least one point ($\forall h$) for which $\theta_{zz}\left(h,\xi\right)=m_{zz}\pi/2$ - at this point $\text{CV}\left(\xi\right)\propto\sin\left(2\theta_{zz}\right)$ vanishes, as required. the LH decomposition is recovered with setting $h=1$.

Clearly, the arguments shown above similarly hold for $\theta_{xx}$ and $\theta_{yy}$. Furthermore, we note that $\text{CV}\left(\xi\right)$ has several roots. We choose the root that corresponds to the lowest value of $|\theta_{xx}|+|\theta_{yy}|+|\theta_{zz}|$, as this will typically correspond to a higher fidelity implementation of $U$.

\begin{figure}[t]
	\includegraphics[width=0.5\textwidth]{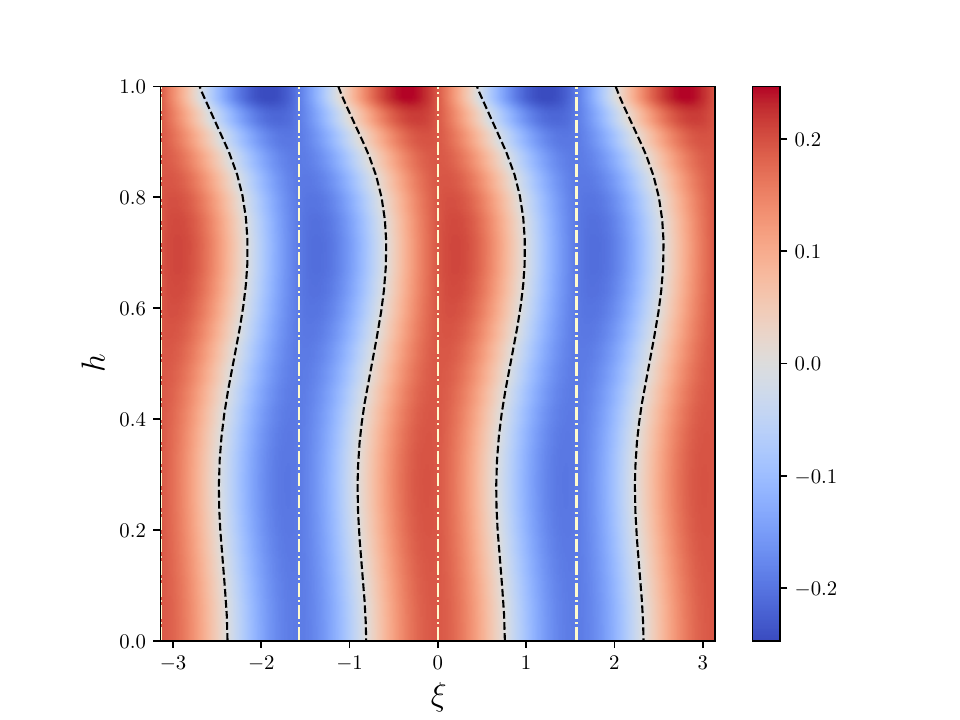}
	\caption{The value of CV (color) of $U_\text{random}$ for varying values of $\xi$ and $h$. The LH decomposition is formed for $\text{CV}=0$ (black dashed line) and varies smoothly with the homotopy from $h=0$ to $h=1$. Lines of constant values of CV are seen for $\xi=0,\pm\pi/2,\pm\pi$ (white dot-dashed lines), as this is equivallent to single qubit rotations of $U_\text{random}$. }
	\label{figLH}
\end{figure}

Figure \ref{figLH} shows an example of the CV (color) as a function of $\xi$ and $h$ for an arbitrary $U_\text{random}\in SU\left(4\right)$. Here specifically we use,
\begin{equation}
	U_\text{random}=\begin{bmatrix}
		0.46 - 0.37i & 0.34 - 0.23i & 0.34 + 0.18i & -0.15 + 0.56i \\
		-0.57 + 0.18i & -0.24 - 0.20i & 0.23 + 0.60i & -0.23 + 0.27i \\
		-0.04 - 0.13i & 0.16 - 0.22i & 0.63 - 0.02i & -0.23 - 0.68i \\
		0.12 - 0.51i & -0.54 + 0.60i & 0.16 + 0.14i & -0.16 - 0.04i
	\end{bmatrix}+\mathcal{O}\left(10^{-3}\right).
\end{equation}
As shown, the value of $\xi$ for which $\text{CV}=0$ and the LH decomposition is formed (black dashed line) varies smoothly with $h$ from the trivial solution of $h=0$ for which $\xi_0=\theta_{zz}$, to the decomposition of $U_\text{random}$ at $h=1$. For the case $\xi=0,\pm\pi/2,\pm\pi$ the rotation by the angle $\xi$ amounts to a single qubit rotation and does not change CV, thus at these values (white dot-dashed lines) the value of CV is fixed.

\subsection{Proof of existence based on algebraic characterization of the Cartan volume}
\label{sec:algebraic_proof}
In this section we provide a simple algebraic expressions for $\text{CV}\left(U\right)$, the Cartan volume of $U$ and compute from it a simple algebraic expression for $v(\xi) \equiv \text{CV}\left(U e^{-i \xi ZZ }\right)$.
As explained in the main text, a LH decomposition of $U$ is easily obtained given a $\xi_0$ zero of $v(\xi)$.

Here and below we follow the Cartan decomposition conventions of the previous section.
Let $U_\text{MB} \equiv M^\dagger U M = O_1^T D O_2 $ with 
$O_1, O_2$ being orthogonal matrices and $D=\text{diag}(\lambda_1, \lambda_2, \lambda_3, \lambda_4)$ being a diagonal matrix with eigenvalues $\lambda_1, \lambda_2, \lambda_3, \lambda_4$ of the form,
\begin{equation}
	\begin{aligned}	
		\lambda_1 &= e^{i(+\theta_{xx} - \theta_{yy} + \theta_{zz})} \\
		\lambda_2 &= e^{i(+\theta_{xx} + \theta_{yy} - \theta_{zz})} \\
		\lambda_3 &= e^{i(-\theta_{xx} - \theta_{yy} - \theta_{zz})} \\
		\lambda_4 &= e^{i(-\theta_{xx} + \theta_{yy} + \theta_{zz})}\label{eqLambda}
	\end{aligned}
\end{equation}

where $\theta_{xx}, \theta_{yy}, \theta_{zz}$ are the Cartan two qubit phases of the corresponding $R_\text{XX}$ ,$R_\text{YY}$ and $R_\text{ZZ}$ rotations.
Noting that, $ \lambda_1 \lambda_2 = e^{i2\theta_{xx}}$,
$ \lambda_1 \lambda_3=e^{-i2\theta_{yy}} $ and $ \lambda_2 \lambda_3=e^{-i2\theta_{zz}} $ we obtain,
\begin{equation}
	\text{CV}\left(U\right) = \text{sin}(2\theta_{xx})\text{sin}(2\theta_{yy})\text{sin}(2\theta_{zz})=\text{Im}(\lambda_1 \lambda_2)\text{Im}(\lambda_1 \lambda_3)\text{Im}(\lambda_2 \lambda_3).
\end{equation}

Then, using the relations, $\lambda_j \lambda^*_j=1 $ and $\lambda_1 \lambda_2 \lambda_3 \lambda_4 = 1$ we obtain,
\begin{equation}
	\text{Im}(\lambda_1 \lambda_2 )\text{Im}(\lambda_2 \lambda_3 )=\frac{1}{4} (\lambda_1 \lambda_2-\lambda_1^* \lambda_2^* )
	(\lambda_2 \lambda_3-\lambda_2^* \lambda_3^* )=\frac{1}{2} \text{Re}(\lambda_2 \lambda_4^*-\lambda_1 \lambda_3^* ).
\end{equation}
Since, $\lambda_1 \lambda_2 \lambda_3=\lambda_4^*$ and similarly $\lambda_4^* \lambda_1^* \lambda_3^*=\lambda_2$ (and similarly for other products of the eigenvalues), it follows that,
\begin{equation}
	\begin{aligned}
		\text{Im}(\lambda_1 \lambda_2 )\text{Im}(\lambda_2 \lambda_3 )\text{Im}(\lambda_1 \lambda_3 )&=\frac{1}{8} (\lambda_2 \lambda_4^*-\lambda_1 \lambda_3^*+\lambda_2^* \lambda_4-\lambda_1^* \lambda_3 )(\lambda_1 \lambda_3-\lambda_1^* \lambda_3^* )\\
		&= \frac{1}{8} ((\lambda_4^* )^2-(\lambda_2 )^2-(\lambda_1 )^2+(\lambda_3^* )^2+(\lambda_2^* )^2-(\lambda_4 )^2-(\lambda_3 )^2+(\lambda_1^* )^2 )\\
		&= -\frac{1}{4} \text{Im}((\lambda_1 )^2+(\lambda_2 )^2+(\lambda_3 )^2+(\lambda_4 )^2 )\\
		&= -\frac{1}{4} \text{Im}(\text{Tr}(F^2)) \\
		&=-\frac{1}{4} \text{Im}(\text{Tr}(O_2^T F^2 O_2))\\
		&=-\frac{1}{4} \text{Im}(\text{Tr}(U_\text{MB}^T U_\text{MB})).
	\end{aligned}
\end{equation}

This yields a short algebraic expression for the Cartan volume of $U$,
\begin{equation}
	\label{eq:CV_formula}
\text{CV}\left(U\right)=\text{sin}(2\theta_{xx})\text{sin}(2\theta_{yy})\text{sin}(2\theta_{zz})
=
-\frac{1}{4}\text{Im}\left( \text{Tr} \left( U_{\text{MB}}^T U_{\text{MB}} \right) \right)
\end{equation}

Now for the LH decomposition, recalling that $M e^{-i\xi ZZ}=e^{-i\xi ZZ}M$, we define,
\begin{equation}
	U_\text{MB} (\xi) = M^\dagger  U e^{-i\xi ZZ} M = M^\dagger U M e^{-i\xi ZZ}=U_\text{MB} e^{-i\xi ZZ}.
\end{equation}

Noting that $U e^{-i \xi ZZ } \in SU(4)$, we compute the Cartan volume $\text{CV}(U e^{-i \xi ZZ })$
with (\ref{eq:CV_formula}) and obtain,
\begin{equation}
	\text{CV}(U e^{-i \xi ZZ })=	-\frac{1}{4} \text{Im}(\text{Tr}(e^{-i\xi ZZ} U_\text{MB}^T U_\text{MB} e^{-i\xi ZZ} ))-\frac{1}{4} \text{Im}\left( (a_r+a_i i)e^{-i2\xi}+(b_r+b_i i)e^{i2\xi} \right),
\end{equation}

with,
\begin{equation}
	\begin{aligned}
		a_r+a_i i \equiv \sum_{j=1)}^4 (U_\text{MB})_{j,1)}^2 +(U_\text{MB})_{j,4)}^2 ) \\ 
		b_r+b_i i \equiv \sum_{j=1)}^4 (U_\text{MB})_{j,2)}^2 +(U_\text{MB})_{j,2)}^2 ).
	\end{aligned}
\end{equation}

Thus, we obtain,
\begin{equation}
v(\xi) \equiv \text{CV}(U e^{-i \xi ZZ })=\frac{1}{4} \left(-(a_i+b_i)\text{cos}(2\xi)+(a_r-b_r)\text{sin}(2\xi) \right) =
 a(U) \text{sin} \left(2(\xi-\xi_0)\right)
\end{equation}
where, 
\begin{align*}
a(U)\equiv \frac{1}{4} \sqrt{(a_i+b_i)^2+(a_r-b_r)^2} \\
\xi_0(U)\equiv\frac{1}{4}
\text{arctan}\left(\frac{a_i+b_i}{a_r-b_r}\right)
\end{align*}
Finally, as $v\left( \xi_0(U) \right) $ vanishes, $\xi_0(U)$ is the required phase of the initial MS gate in the LH-decomposition.

\subsection{LH compilation entanglement phase optimization}

We describe minimization of the entanglement phases of the LH decomposition using single qubit $R_\text{Y}$ rotations. Consider the LH decomposition of a generic $SU4$ gate denoted by $G$,
\begin{equation}
	G = \left(\tilde{Q}_0 \otimes \tilde{Q}_1 \right) \cdot e^{-i\frac{\pi}{4} \tilde{\gamma} ZZ} \cdot e^{-i\frac{\pi}{4} \tilde{\beta} XX} 
	\left(\tilde{P}_0 \otimes \tilde{P}_1 \right) \cdot e^{-i\frac{\pi}{4} \tilde{\alpha} ZZ}.
\end{equation}
and consider also the Cartan decomposition of the same $SU4$ gate $G$,
\begin{equation}
	G = \left(Q_0 \otimes Q_1 \right) \cdot e^{-i\frac{\pi}{4} {\gamma} ZZ} \cdot e^{-i\frac{\pi}{4} {\beta} YY}
	\cdot e^{-i\frac{\pi}{4} {\alpha} XX}\cdot \left({P}_0 \otimes {P}_1 \right).
\end{equation}
where the entangling phases are chosen to minimize the nuclear norm, so that $ -1 < \alpha, \beta, \gamma, \tilde{\alpha}, \tilde{\beta}, \tilde{\gamma} \le 1 $.

It turns out that the $L_1$-norm of LH decompositions $|\tilde{\alpha}|+|\tilde{\beta}|+|\tilde{\gamma}|$ is always larger or equal when compared with that of the Cartan decomposition $|{\alpha}|+|{\beta}|+|{\gamma}|$.
That is,
\begin{equation}
	|\tilde{\alpha}|+|\tilde{\beta}|+|\tilde{\gamma}| \ge |{\alpha}|+|{\beta}|+|{\gamma}|.
\end{equation}
Importantly, it turns out that the $SU(4)$ operator, $G$, can be modified with two additional initial $R_\text{Y}$ rotations, chosen so that the LH decomposition of the modified gate $\tilde{G} = G\cdot \left( e^{-i\frac{\pi}{2}y_0}\otimes e^{-i\frac{\pi}{2}y_1} \right)$
is such that it has a reduced $L_1$-norm. Moreover, it turns out that there are always phase values $y_0$ and $y_1$ such that the $L_1$-norm of $\tilde{G}$ becomes optimal and equal to the $L_1$-norm of the Cartan decomposition $|{\alpha}|+|{\beta}|+|{\gamma}|$ of $G$ and $\tilde{G}$.

\begin{figure}[H]
	\includegraphics[width=0.5\textwidth]{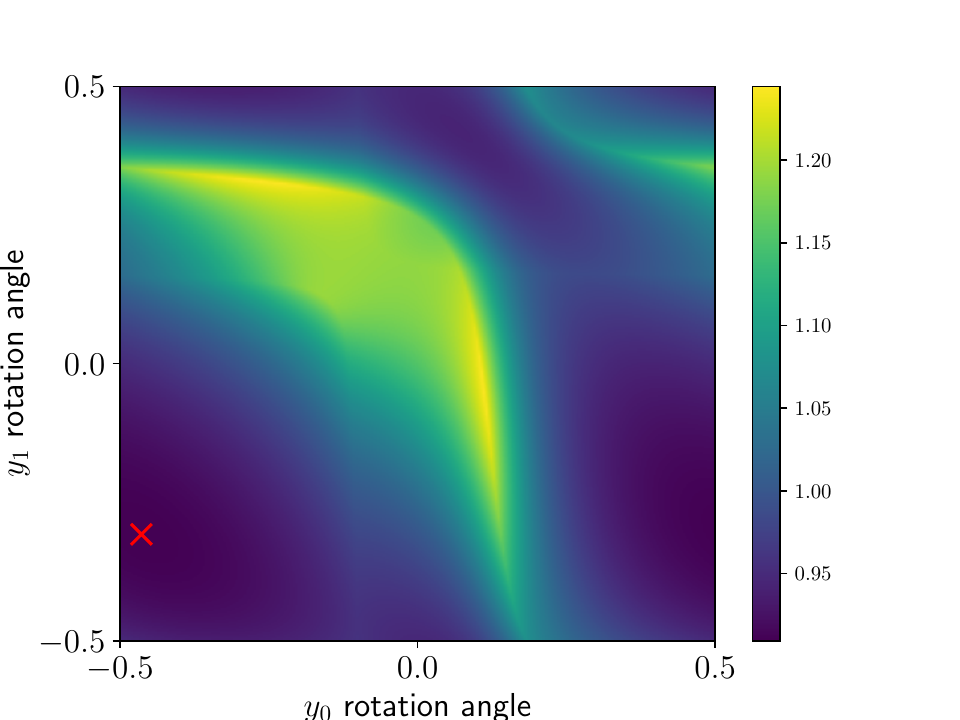}
	\caption{The $L_1$-norm optimization landscape of the same random unitary, $U_\text{random}$ shown in Fig. \ref{figLH}, different points correspond to deformations of $U_\text{random}$ obtained by preceding the gate with single qubit $R_\text{Y}$ rotations with angles $y_0$ and $y_1$. The  $L_1$-norm of the resulting LH decomposition is shown (color) with the minimal obtainable value marked by a red cross.}
	\label{figRyLandscape}
\end{figure}

Figure \ref{figRyLandscape} shows the $L_1$-norm optimization landscape of the same random unitary shown in Fig. \ref{figLH}, as a function of $y_0$ and $y_1$. Clearly, various choices of rotation angles can effect the $L_1$-norm by more than $20\%$. The optimal (minimal) choice is marked by a red cross.

\subsection{`Pull-back' of Y gate rotations}\label{sec:pull-back-scheme}

Let $G$ be a LH decomposition of the form,
\begin{equation}
	G =  \left(e^{-i\frac{\pi}{2}y_0 Y_0} \otimes e^{-i\frac{\pi}{2}y_1 Y_1} \right) \cdot e^{-i\frac{\pi}{4} \gamma ZZ} \cdot e^{-i\frac{\pi}{4} \beta XX} \left(P_0 \otimes P_1 \right) \cdot e^{-i\frac{\pi}{4} \alpha ZZ}.
	\label{LH-decom-Y_gates}
\end{equation}
In the following we show a compilation technique that we use to `pull-back' the final two $R_\text{Y}$ rotations, merging them into the LH decomposition, making this decomposition both left hand and right-hand (RH) decomposition, i.e. starting and ending with $R_\text{ZZ}$ rotations. Crucially, this technique provides two additional degrees of freedom, $y_0$ and $y_1$, that we use for circuit nuclear norm optimization.

Specifically, given a LH-decomposition as in \eqref{LH-decom-Y_gates}, we rearrange the layers of $G$ to a LH-RH decomposition form, as follows,
\begin{equation}
	G =  e^{-i\frac{\pi}{4} \tilde{\gamma} ZZ} Z_0^{\tilde{\ell}_z} \otimes Z_1^{\tilde{\ell}_z}
	e^{ -i\frac{\pi}{2} \tilde{u}_0 Y_0 } \otimes e^{-i\frac{\pi}{2} \tilde{u}_1 Y_1 }  e^{-i\frac{\pi}{4} \tilde{\beta} XX} 
	X_0^{\tilde{\ell}_x} \otimes X_1^{\tilde{\ell}_x} 
	e^{-i\frac{\pi}{2} \tilde{v}_0 Y_0}P_0 \otimes e^{-i\frac{\pi}{2}\tilde{v}_1 Y_1} P_1 f	e^{-i\frac{\pi}{4} \alpha ZZ},
\end{equation}
with the phase parameters, $ -1 < \tilde{\beta}, \tilde{\gamma},\tilde{\ell}_0, \tilde{u}_0, \tilde{u}_1, \tilde{v}_0, \tilde{v}_1 < 1 $ and $\tilde{\ell}_x,\tilde{\ell}_z\in\{0,1\}$ 

These are obtained using single qubit Euler decomposition \cite{crooks2024quantum}, as follows:
\begin{enumerate}
	\item With Euler decomposition let us find $ p_x, p_y $ and $ p_z $ be such that, $e^{-i\frac{\pi}{2} p_z Z} e^{-i\frac{\pi}{2} p_y Y} e^{-i\frac{\pi}{4} p_x X} = e^{i\frac{\pi}{2}y_0 Y} e^{i\frac{\pi}{4}\alpha Z}$.
	\item With Euler decomposition let us find $ q_1, q_z $ and $ q_2 $ be such that,
	$ e^{-i\frac{\pi}{2} q_{2} Y} e^{-i\frac{\pi}{2} q_z Z} e^{-i\frac{\pi}{4} q_{1} Y} = e^{-i\frac{\pi}{2}p_x X} e^{i\frac{\pi}{4}\beta Z} $.
	\item With Euler decomposition let us find $ u, v $ and $ w $ be such that,
	$ e^{-i\frac{\pi}{2} w Z} e^{-i\frac{\pi}{2} v Y} e^{-i\frac{\pi}{4} u X} = e^{i\frac{\pi}{2}y_1 Y} e^{i\frac{\pi}{4} q_1 Z} e^{i\frac{\pi}{4}  Y}$.
	\item With Euler decomposition let us find $ a, b $ and $ c $ be such that,
	$ e^{-i\frac{\pi}{2} a Y} e^{-i\frac{\pi}{2} b Z} e^{-i\frac{\pi}{4} c Y} = e^{-i\frac{\pi}{2}w X} e^{i\frac{\pi}{4} q_1 Y} e^{i\frac{\pi}{4}p_z  Z}$.
	\item Finally, set  $ \tilde{u}_0 \equiv a$,  $ \tilde{u}_1 \equiv -v$,
	$ \tilde{v}_0 \equiv -c $,
	$ \tilde{v}_1 \equiv -q_z $,
	$ \tilde{\ell}_x \equiv \text{round}(-b) $, $ \tilde{\ell}_z \equiv \text{round}(-u) $,
	$\tilde{\beta} \equiv -b-\text{round}(-b) $ and
	$\tilde{\gamma} \equiv -u-\text{round}(-u) $
	with $ \text{round}(x) $ denoting the integer which is closest to $x$.
\end{enumerate}

\section{Nuclear norm Optimization algorithm}
\begin{figure}
	
	\begin{tabular}{cc}
		{\bf (a)} & 
		{\bf (b)} \\
		\includegraphics[width=0.45\textwidth]{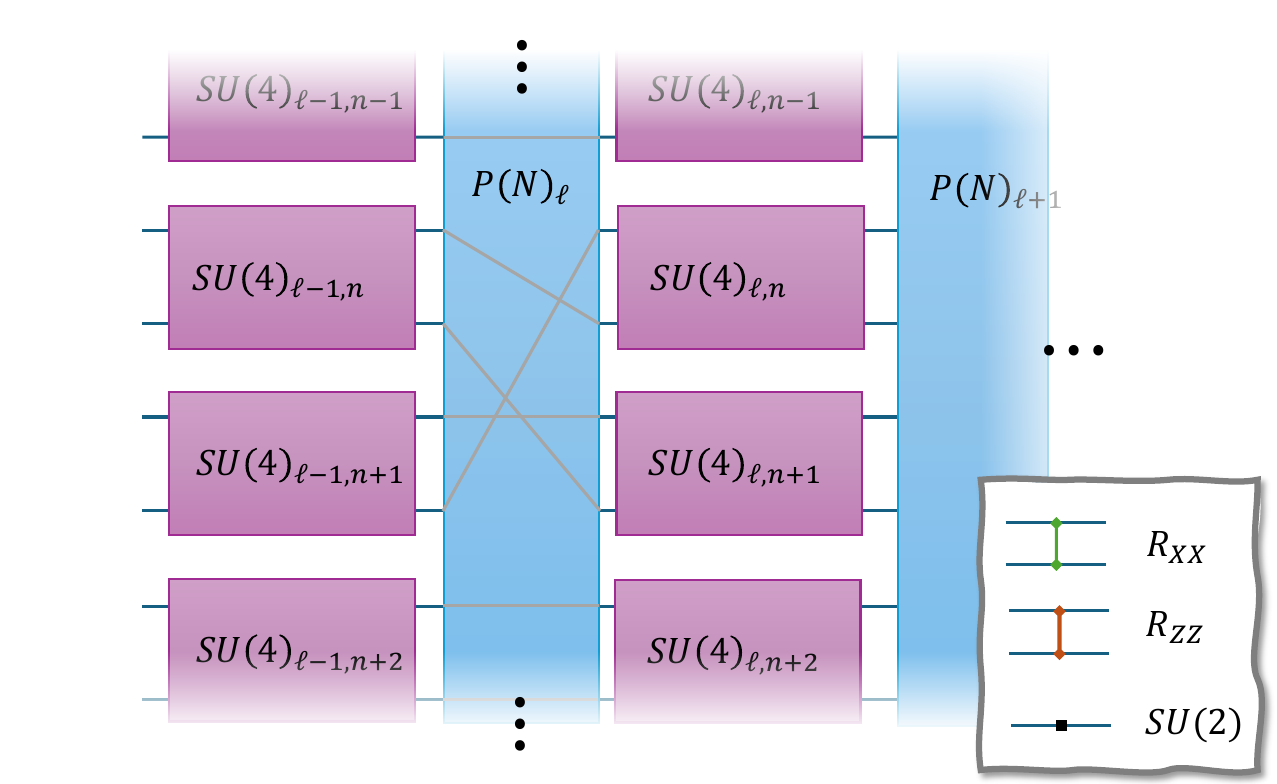} &
		\includegraphics[width=0.45\textwidth]{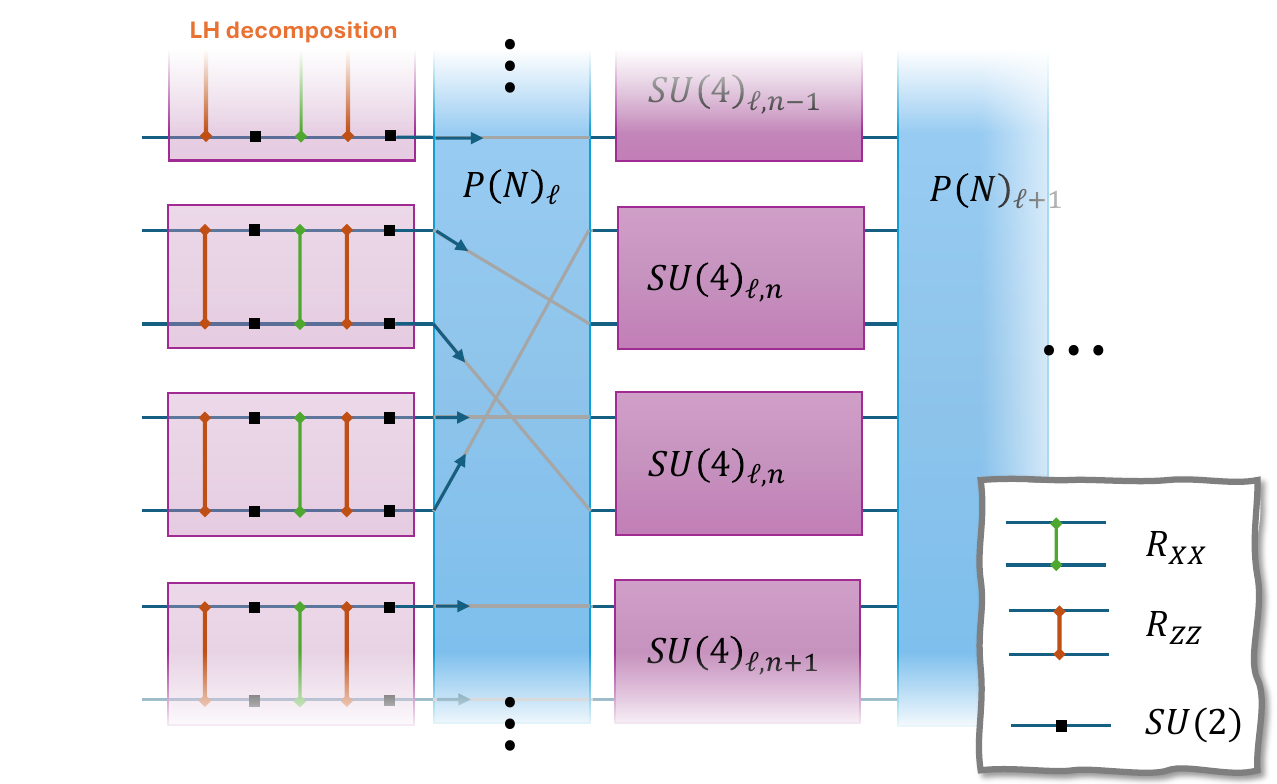} \\

		{\bf (c)}
		& 
		{\bf (d)} \\
		\includegraphics[width=0.45\textwidth]{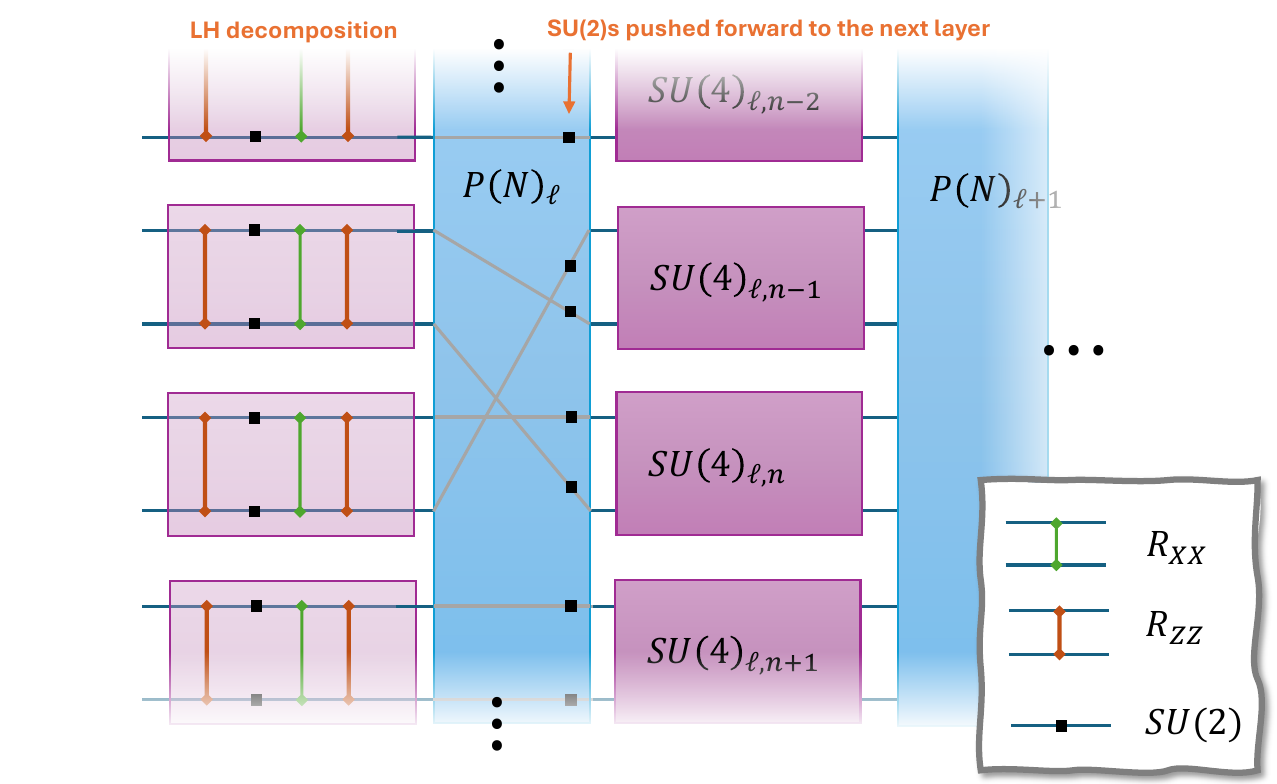} &
		\includegraphics[width=0.45\textwidth]{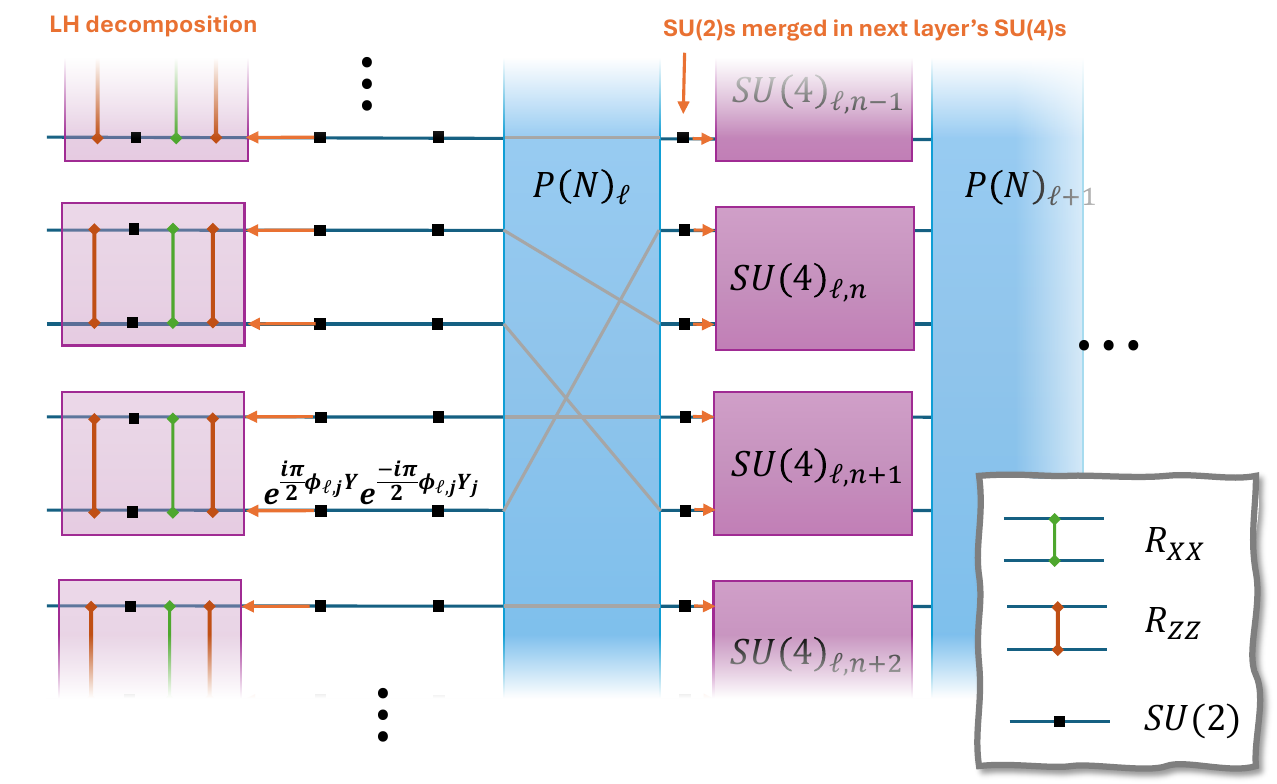} \\

		{\bf (e)} & {\bf (f)} \\	
		\includegraphics[width=0.45\textwidth]{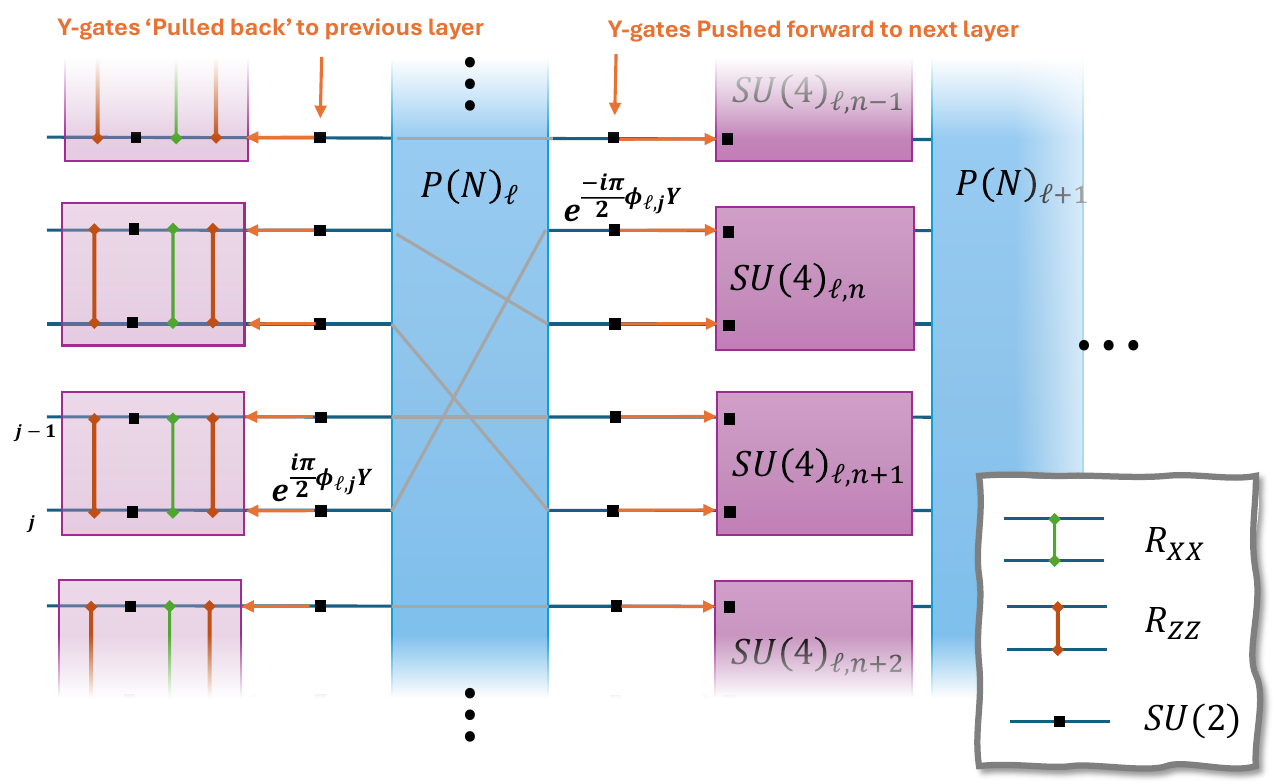} &
		\includegraphics[width=0.45\textwidth]{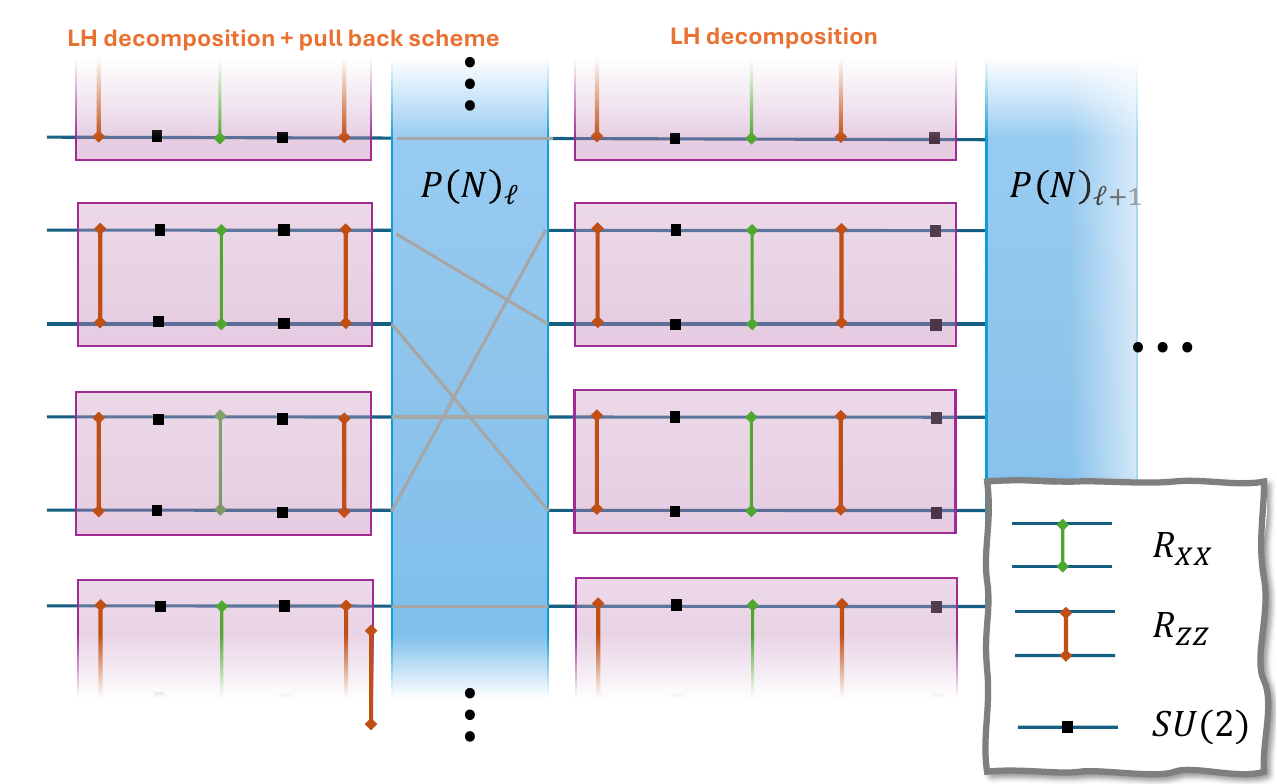} \\

	\end{tabular}
	\caption{Nuclear norm optimization algorithm steps. (a) Initial QV circuit. (b) LH decompositions of the $\ell^{\text th}$ layer's $SU(4)$ gates, with  $(\ell-1)^{\text th}$ layer already LH-decomposed. (c) Last $SU(2)$ gates of the $(\ell-1)^{\text th}$ layer's LH-decompositions 'pushed-forward' to the $\ell^{\text th}$ layer and merged in it's $SU(4)$ gates. (d) Two additional layers of $Y$-gate rotations $e^{\frac{i\pi}{2}\sum_{j=1}^{n_q}\phi_{\ell,j}Y_j}$ and $e^{\frac{-i\pi}{2}\sum_{j=1}^{n_q}\phi_{\ell,j}Y_j}$ added to the circuit (the subscript $j$ is a qubit number index). The phases $\phi_{\ell,j}$ of these gates are chosen to optimize the combined nuclear norm of the subsequent (e-f) compilation steps of $\ell^{\text th}$ and $(\ell+1)^{\text th}$ layers. (e) The $e^{\frac{i\pi}{2}\sum_{j=1}^{n_q}\phi_{\ell,j}Y_j}$ $Y$-gates layer is re-compiled with the 'pull-back' scheme in the $\ell^{\text th}$ layer and the $e^{\frac{-i\pi}{2}\sum_{j=1}^{n_q}\phi_{\ell,j}Y_j}$, $Y$-gates layer is 'pushed-forward' and merged in the $SU(4)$s of the $(\ell+1)^{\text th}$ layer. 	(f) Finally, the (newly modified) $SU(4)$ gates of the $(\ell+1)^{\text th}$ layer are compiled with the LH-decomposition scheme - So that the last $zz$ gates of the $(\ell)^{\text th}$ layer and the first $zz$ gates of the $(\ell+1)^{\text th}$ layer are merged into a single MQ layer. Then these (a-f) are iterated in the next layers, until the whole circuit is recompiled.}\label{fig:Algorithm}
\end{figure}

We describe the layer-by-layer optimization algorithm that minimizes the total nuclear norm of the compiled circuit. Figure \ref{fig:Algorithm} presents this procedure graphically and Algorithm \ref{alg:optimization_algo} describe this in pseudo-code.

\subsection{Definitions and notations used within the pseudo code}
\begin{itemize}
	\item \textbf{LH decomposition:}\newline
	Denote the `LH decomposition' of a two qubit $U_{(j_1,j_2)}\in SU(4)$ gate operating on a qubit pair $(j_1,j_2)$, with $\text{LH}\left[ U_{(j_1,j_2)} \right]$. This notation indicates a compilation scheme of $U_{(j_1,j_2)}$, having the following form,
	\begin{equation}
			\text{LH}\left[ U_{(j_1,j_2)} \right] \equiv \left( A_{j_1}\otimes A_{j_2}\right) e^{\frac{-i\pi}{4}\gamma Z_{j_1}Z_{j_2}} e^{\frac{-i\pi}{4}\beta X_{j_1}X_{j_2}}	\left( B_{j_1}\otimes B_{j_2}\right) 	e^{\frac{-i\pi}{4}\alpha Z_{j_1}Z_{j_2}},
	\end{equation}
	where $A_{j_1}, A_{j_2}, B_{j_1}, B_{j_2}$ are $SU(2)$ gates and $ -1 < \alpha, \beta, \gamma \le 1 $ are phases of correlated qubit rotations.
	To define this compilation scheme uniquely, we choose the compilation that has the smallest possible $L_1$-norm
	and satisfies the condition: $|\beta| \le \text{min}(|\alpha|,|\gamma|) $.
	\item \textbf{LH-bare decomposition  and  push-forward gates associated with a LH decomposition:}
	\newline
	Given a LH decomposition, $\text{LH}\left[ U_{(j_1,j_2)} \right]$, as above, define the `LH bare decomposition' of $U_{(j_1,j_2)}$ as,
	\begin{equation}
			\text{LH}_{\text{bare}}
		\left[ U_{(j_1,j_2)} \right] \equiv
		e^{\frac{-i\pi}{4}\gamma Z_{j_1}Z_{j_2}}
		e^{\frac{-i\pi}{4}\beta X_{j_1}X_{j_2}}
		\left( B_{j_1}\otimes B_{j_2}\right)
		e^{\frac{-i\pi}{4}\alpha Z_{j_1}Z_{j_2}},
	\end{equation}
	and define the corresponding `push-forward' gates as,
	\begin{equation}
			\text{A}_{\text{push-forward}} \left[ U_{(j_1,j_2)} \right]
		\equiv A_{j_1}\otimes A_{j_2}.
	\end{equation}

	Note that, $\text{LH}_{\text{bare}} \left[ U_{(j_1,j_2)} \right] \ne U_{(j_1,j_2)}$.
	To realize equality, we should multiply the left hand side (of the previous inequality) with the single qubit gates $\left( A_{j_1}\otimes A_{j_2}\right)$ - i.e.,
	\begin{equation}
			\text{A}_{\text{push-forward}} \left[ U_{(j_1,j_2)} \right]  \text{LH}_{\text{bare}} \left[ U_{(j_1,j_2)} \right] = U_{(j_1,j_2)}
	\end{equation}

	\item \textbf{Pull-back of $R_\text{Y}$-gates into a LH bare decomposition:}\newline
	When a LH bare decomposition, $\text{LH}_{\text{bare}} \left[ U_{(j_1,j_2)} \right]$, is multiplied from the left with two single qubit $R_\text{Y}$-gate rotations, $e^{\frac{-i\pi}{2}(\theta_{j_1}Y_{j_1}+\theta_{j_2}Y_{j_2})}$,
	define the `Y-pull-back' compilation of this circuit,
	\begin{equation}
		\begin{aligned}
			\text{Y}_\text{pull-back} & \left[ e^{\frac{-i\pi}{2}(\theta_{j_1}Y_{j_1}+\theta_{j_2}Y_{j_2})}\text{LH}_{\text{bare}}
			\left[ U_{(j_1,j_2)} \right] \right] \equiv \\
			& e^{\frac{-i\pi}{4}\tilde{\gamma} Z_{j_1}Z_{j_2}}
			\left( \tilde{A}_{j_1}\otimes \tilde{A}_{j_2}\right)
			e^{\frac{-i\pi}{4}\tilde{\beta} X_{j_1}X_{j_2}}
			\left( \tilde{B}_{j_1}\otimes \tilde{B}_{j_2}\right)
			e^{\frac{-i\pi}{4}{\alpha} Z_{j_1}Z_{j_2}}
		\end{aligned}
	\end{equation}
	The details of the Y-pull-back scheme (including formulas for $\tilde{\beta},\tilde{\gamma},\tilde{A}_{j_1},\tilde{A}_{j_2},\tilde{B}_{j_1},\tilde{B}_{j_2}$) are given in section-\ref{sec:pull-back-scheme}.

	\item \textbf{Definition of the product
		${ \prod_{\ell=1}^{L} A_\ell }$
		of non-commuting operator ${ A_1,A_2,...,A_L}$:}
	\newline
	In this work we use the ordered notation:
	\[\prod_{\ell=1}^{L} A_\ell \equiv A_L  A_{L-1}  A_{L-2} \cdots A_1 \]
\end{itemize}

\subsection{Optimization algorithm pseudo-code}
\begin{itemize}
	\item \textbf{Input:} A quantum volume circuit $\text{QV}(L,N)$ 
	operating on $N$ qubits, with $L$ layers, such that
	each layer is composed of 
	a product of ${\lfloor N/2 \rfloor}$ 
	disjoint  $U_{\ell,\left(j_1(\ell,n),j_2(\ell,n)\right)}\in SU(4)$ gates, such that,
	\begin{equation}
		\text{QV}(L,N)=\prod_{\ell=1}^L {\left( \left[\prod_{n=1}^{\lfloor N/2 \rfloor} 
			{U_{\ell,\left(j_1(\ell,n),j_2(\ell,n)\right)}}\right]  \right) },
	\end{equation}
	where, for every $\ell$, all the two qubit operators $\{U_{\ell,\left(j_1(n),j_2(n)\right)}\}_{n=1}^{N/2}$ 
	operate on different disjoint pairs of qubit - so that, whenever $n \ne m$
	$ \{j_1(\ell,n),j_2(\ell,n)\} \cap \{j_1(\ell,m),j_2(\ell,m)\} = \varnothing . $
	
	\item \textbf{Output:} 
	A compiled circuit $\text{CIRC}_{\text{MQ}}[\text{QV}(L,N)]$ containing $2L+1$ layers of sequential $R_\text{ZZ}$ gates that are trivially merged to MQ gates, such that they optimized to obtain a reduced total nuclear norm.
	The form of this new compilation is as follows,
	\begin{equation}
		\text{CIRC}_{\text{MQ}}[\text{QV}(L,N)] =
		A(\ell=L)\left[\prod_{n=1}^{\lfloor N/2 \rfloor} \text{LH}_\text{bare}\left[{U_{\ell,\left(j_1(\ell,n),j_2(\ell,n)\right)}^\prime}\right]\right]
		 B(\ell=0)
	\end{equation}

	where, $U_{\ell,\left(j_1(\ell,n),j_2(\ell,n)\right)}^\prime$ are a set of modified $SU(4)$ operators that are derived during the optimization from the original QV circuit and $A(\ell=L)$ and $B(\ell=0)$ are layers of single qubit, $SU(2)$, operators acting on each qubit.
	
\end{itemize}

\begin{algorithm}
	\caption{Optimized compilation aimed to minimize both nuclear norm and number of MQ layers}

	\label{alg:optimization_algo}
	
	\begin{algorithmic}[1]
		\Statex
		\Statex
		\Require
		A quantum volume circuit $\text{QV}(L,N)$ operating on $N$ qubits,
		with $L$ layers,
		\[ \text{QV}(L,N)=\prod_{\ell=1}^L {\left( \left[\prod_{n=1}^{\lfloor N/2 \rfloor} {U_{\ell,\left(j_1(\ell,n),j_2(\ell,n)\right)}}\right]  \right) }. \]
		\Statex
		\Ensure
		A compiled circuit $\text{CIRC}_{\text{MQ}}[\text{QV}(L,N)]$ with $(2L+1)$ MQ layers and minimized total nuclear norm.
		\Statex
		\State
		{\bf Copy} Unitary operators $\{U_{\ell,\left(j_1(\ell,n),j_2(\ell,n)\right)}^\prime\}_{\ell=1,...L;n=1,...,\lfloor N/2 \rfloor}=\{U_{\ell,\left(j_1(\ell,n),j_2(\ell,n)\right)}\}_{\ell=1,...L;n=1,...,\lfloor N/2 \rfloor}$
		\Statex
		\State
		{\bf Define} phases $\psi_{\ell,j}$, $j=1,2,3,...,N$,
		\[\{ \psi_j \}_{j=1}^{N} =\operatorname*{arg\,min}_{\phi_j} \left\lVert \prod_{n=1}^{\lfloor N/2 \rfloor} \text{LH} \left[ U_{\ell=1,\left(j_1(\ell=1,n),j_2(\ell=1,n)\right)}^\prime
		e^{ \frac{-i\pi}{2} \left( \phi_{j_1(\ell=1,n)} Y_{j_1(\ell=1,n)} +	\frac{-i\pi}{2} \phi_{j_2(\ell=1,n)} Y_{j_2(\ell=1,n)}\right)} \right] \right\rVert_{L_1}.\]
		\Statex
		\State
		{\bf Define} the $\ell=0$ single-qubit layer of the compiled circuit, $ B[\ell=0] =
		e^{\frac{i\pi}{2}\sum_{n=1}^{\lfloor N/2 \rfloor}\left(\psi_{j_1(\ell=1,n)}Y_{j_2(\ell=1,n)}  + \psi_{j_2(\ell=1,n)}Y_{j_2(\ell=1,n)}\right) }$
		\Statex
		\State
		{\bf Update} the first $\ell=1$ layer of the QV-circuit according to
		\[ \prod_{n=1}^{\lfloor N/2 \rfloor}
		U_{\ell,\left(j_1(\ell,n),j_2(\ell,n)\right)}^\prime
		\rightarrow  \prod_{n=1}^{\lfloor N/2 \rfloor}\left[ U_{\ell,\left(j_1(\ell,n),j_2(\ell,n)\right)}^\prime
		e^{			 
		\frac{-i\pi}{2} \left(\phi_{j_1(\ell,n)} Y_{j_1(\ell,n)} +
		\phi_{j_2(\ell,n)} Y_{j_2(\ell,n)}
		\right)}\right]
		\]
		\For{$ \ell = 1,2,3,...,(L-1) $}
		\Statex
		\State 
		{\bf Define} phases $\phi_{\ell,j}$, $j=1,2,3,...,N$,
		\begin{align*}
			\{ \phi_{\ell,j} \}_{j=1}^{N} = 
			\operatorname*{arg\,min}_{\phi_{\ell,j}}
			& \left\lVert
			\text{LH}
			\left[ \prod_{n=1}^{\lfloor N/2 \rfloor} 
			\left(
			U_{ \ell+1,\left(j_1(\ell+1,n),j_2(\ell+1,n)\right) }^\prime \right)
			\prod_{n'=1}^{\lfloor N/2 \rfloor} \text{A}_{\text{push-forward}} 
			\left[
			U_{\ell,\left(j_1(\ell,n'),j_2(\ell,n')\right)}^\prime \right]\right]\right.
			\\
			& \left. \prod_{n'=1}^{\lfloor N/2 \rfloor} \text{Y}_\text{pull-back} \left[  
			e^{ \frac{i\pi}{2} 
				\left( \phi_{\ell,j_1(\ell,n')}Y_{j_1(\ell,n')} + \phi_{\ell,j_2(\ell,n')}Y_{j_2(\ell,n')} \right) } \text{LH}_{\text{bare}}\left[ 
			U_{\ell,\left(j_1(\ell,n'),j_2(\ell,n')\right)}^\prime \right]   \right]
			\right\rVert_{L_1}.
		\end{align*}
		\State
		{\bf Define} the $\ell$-th layer of the compiled circuit to be,
		\[ \text{CIRC}_{\text{MQ}}[\ell] =
		\prod_{n'=1}^{\lfloor N/2 \rfloor} \text{Y}_\text{pull-back} \left[  
		e^{ \frac{i\pi}{2} 
			\left( \phi_{\ell,j_1(\ell,n)}Y_{j_1(\ell,n)} + \phi_{\ell,j_2(\ell,n)}Y_{j_2(\ell,n)} \right) } \text{LH}_{\text{bare}}\left[ 
		U_{\ell,\left(j_1(\ell,n),j_2(\ell,n)\right)}^\prime \right]   \right]
		\]
		\State
		{\bf Update} the $(\ell+1)$-th layer of the QV-circuit with single qubit gates, pushed forward from the previous layer,
		\[ \prod_{n=1}^{\lfloor N/2 \rfloor}U_{\ell+1,n}
		\rightarrow   
		\left[\prod_{n=1}^{\lfloor N/2 \rfloor}
		\left(U_{ \ell+1,\left(j_1(\ell+1,n),j_2(\ell+1,n)\right) }^\prime \right)
		\prod_{n'=1}^{\lfloor N/2 \rfloor} \text{A}_{\text{push-forward}} 
		\left[
		U_{\ell,\left(j_1(\ell,n'),j_2(\ell,n')\right)}^\prime \right]\right]
		\]
		\EndFor
		\Statex
		\State
		{\bf Define} the $L$-th layer of the compiled circuit,
		\[ 
		\text{CIRC}_{\text{MQ}}[L] =
		\text{LH}\left[ 
		\prod_{n=1}^{\lfloor N/2 \rfloor} \left(U_{ L,\left(j_1(L,n),j_2(L,n)\right) }^\prime \right)
		\prod_{n'=1}^{\lfloor N/2 \rfloor} \text{A}_{\text{push-forward}} 
		\left[
		U_{L-1,\left(j_1(L-1,n'),j_2(L-1,n')\right)}^\prime \right] 
		\right] 
		\]
		\State
		{\bf Define} the compiled MQ nuclear norm optimized circuit as,
		\[ \text{CIRC}_{\text{MQ}} = \prod_{\ell=1}^L \text{CIRC}_{\text{MQ}}[\ell]B(\ell=0)  \]
	\end{algorithmic}
\end{algorithm}
\newpage

%% file: manuscript.bbl
\begin{thebibliography}{10}

\bibitem{shaydulin2023qaoa}
Ruslan Shaydulin and Marco Pistoia.
\newblock Qaoa with $n\cdot p\geq 200$, 2023.

\bibitem{willsch2024state}
Dennis Willsch, Philipp Hanussek, Georg Hoever, Madita Willsch, Fengping Jin,
  Hans~De Raedt, and Kristel Michielsen.
\newblock The state of factoring on quantum computers, 2024.

\bibitem{decross2024computational}
Matthew DeCross, Reza Haghshenas, Minzhao Liu, Enrico Rinaldi, Johnnie Gray,
  Yuri Alexeev, Charles~H. Baldwin, John~P. Bartolotta, Matthew Bohn, Eli
  Chertkov, Julia Cline, Jonhas Colina, Davide DelVento, Joan~M. Dreiling,
  Cameron Foltz, John~P. Gaebler, Thomas~M. Gatterman, Christopher~N. Gilbreth,
  Joshua Giles, Dan Gresh, Alex Hall, Aaron Hankin, Azure Hansen, Nathan
  Hewitt, Ian Hoffman, Craig Holliman, Ross~B. Hutson, Trent Jacobs, Jacob
  Johansen, Patricia~J. Lee, Elliot Lehman, Dominic Lucchetti, Danylo Lykov,
  Ivaylo~S. Madjarov, Brian Mathewson, Karl Mayer, Michael Mills, Pradeep
  Niroula, Juan~M. Pino, Conrad Roman, Michael Schecter, Peter~E. Siegfried,
  Bruce~G. Tiemann, Curtis Volin, James Walker, Ruslan Shaydulin, Marco
  Pistoia, Steven.~A. Moses, David Hayes, Brian Neyenhuis, Russell~P. Stutz,
  and Michael Foss-Feig.
\newblock The computational power of random quantum circuits in arbitrary
  geometries, 2024.

\bibitem{pokharel2023demonstration}
Bibek Pokharel and Daniel~A. Lidar.
\newblock Demonstration of algorithmic quantum speedup.
\newblock {\em Phys. Rev. Lett.}, 130:210602, May 2023.

\bibitem{chowdhury2024enhancing}
Talal~Ahmed Chowdhury, Kwangmin Yu, Mahmud~Ashraf Shamim, M.~L. Kabir, and
  Raza~Sabbir Sufian.
\newblock Enhancing quantum utility: simulating large-scale quantum spin chains
  on superconducting quantum computers, 2024.

\bibitem{shor1994algorithms}
P.W. Shor.
\newblock Algorithms for quantum computation: discrete logarithms and
  factoring.
\newblock In {\em Proceedings 35th Annual Symposium on Foundations of Computer
  Science}, pages 124--134, 1994.

\bibitem{grover1996fast}
Lov~K. Grover.
\newblock A fast quantum mechanical algorithm for database search.
\newblock In {\em Proceedings of the Twenty-Eighth Annual ACM Symposium on
  Theory of Computing}, STOC '96, page 212–219, New York, NY, USA, 1996.
  Association for Computing Machinery.

\bibitem{preskill2018quantum}
John Preskill.
\newblock Quantum {C}omputing in the {NISQ} era and beyond.
\newblock {\em {Quantum}}, 2:79, August 2018.

\bibitem{schwerdt2022comparing}
David Schwerdt, Yotam Shapira, Tom Manovitz, and Roee Ozeri.
\newblock Comparing two-qubit and multiqubit gates within the toric code.
\newblock {\em Phys. Rev. A}, 105:022612, Feb 2022.

\bibitem{goldfriend2024design}
Tomer Goldfriend, Israel Reichental, Amir Naveh, Lior Gazit, Nadav Yoran, Ravid
  Alon, Shmuel Ur, Shahak Lahav, Eyal Cornfeld, Avi Elazari, Peleg Emanuel, Dor
  Harpaz, Tal Michaeli, Nati Erez, Lior Preminger, Roman Shapira, Erik~Michael
  Garcell, Or~Samimi, Sara Kisch, Gil Hallel, Gilad Kishony, Vincent van
  Wingerden, Nathaniel~A. Rosenbloom, Ori Opher, Matan Vax, Ariel Smoler, Tamuz
  Danzig, Eden Schirman, Guy Sella, Ron Cohen, Roi Garfunkel, Tali Cohn, Hanan
  Rosemarin, Ron Hass, Klem Jankiewicz, Karam Gharra, Ori Roth, Barak Azar,
  Shahaf Asban, Natalia Linkov, Dror Segman, Ohad Sahar, Niv Davidson, Nir
  Minerbi, and Yehuda Naveh.
\newblock Design and synthesis of scalable quantum programs, 2024.

\bibitem{heyfron2018efficient}
Luke Heyfron and Earl~T. Campbell.
\newblock An efficient quantum compiler that reduces $t$ count, 2018.

\bibitem{kissinger2020reducing}
Aleks Kissinger and John van~de Wetering.
\newblock Reducing the number of non-clifford gates in quantum circuits.
\newblock {\em Phys. Rev. A}, 102:022406, Aug 2020.

\bibitem{amy2014meet}
Matthew Amy, Dmitri Maslov, Michele Mosca, and Martin Roetteler.
\newblock A meet-in-the-middle algorithm for fast synthesis of depth-optimal
  quantum circuits.
\newblock {\em IEEE Transactions on Computer-Aided Design of Integrated
  Circuits and Systems}, 32(6):818--830, 2013.

\bibitem{debeaudrap2020techniques}
Niel de~Beaudrap, Xiaoning Bian, and Quanlong Wang.
\newblock Techniques to reduce pi/4-parity-phase circuits, motivated by the zx
  calculus.
\newblock {\em Electronic Proceedings in Theoretical Computer Science},
  318:131–149, May 2020.

\bibitem{knill2004fault}
E.~Knill.
\newblock Fault-tolerant postselected quantum computation: Schemes, 2004.

\bibitem{bravyi2005universal}
Sergey Bravyi and Alexei Kitaev.
\newblock Universal quantum computation with ideal clifford gates and noisy
  ancillas.
\newblock {\em Phys. Rev. A}, 71:022316, Feb 2005.

\bibitem{rodriguez2024experimental}
Pedro~Sales Rodriguez, John~M. Robinson, Paul~Niklas Jepsen, Zhiyang He, Casey
  Duckering, Chen Zhao, Kai-Hsin Wu, Joseph Campo, Kevin Bagnall, Minho Kwon,
  Thomas Karolyshyn, Phillip Weinberg, Madelyn Cain, Simon~J. Evered,
  Alexandra~A. Geim, Marcin Kalinowski, Sophie~H. Li, Tom Manovitz, Jesse
  Amato-Grill, James~I. Basham, Liane Bernstein, Boris Braverman, Alexei
  Bylinskii, Adam Choukri, Robert DeAngelo, Fang Fang, Connor Fieweger, Paige
  Frederick, David Haines, Majd Hamdan, Julian Hammett, Ning Hsu, Ming-Guang
  Hu, Florian Huber, Ningyuan Jia, Dhruv Kedar, Milan Kornjača, Fangli Liu,
  John Long, Jonathan Lopatin, Pedro L.~S. Lopes, Xiu-Zhe Luo, Tommaso Macrì,
  Ognjen Marković, Luis~A. Martínez-Martínez, Xianmei Meng, Stefan
  Ostermann, Evgeny Ostroumov, David Paquette, Zexuan Qiang, Vadim Shofman,
  Anshuman Singh, Manuj Singh, Nandan Sinha, Henry Thoreen, Noel Wan, Yiping
  Wang, Daniel Waxman-Lenz, Tak Wong, Jonathan Wurtz, Andrii Zhdanov, Laurent
  Zheng, Markus Greiner, Alexander Keesling, Nathan Gemelke, Vladan Vuletić,
  Takuya Kitagawa, Sheng-Tao Wang, Dolev Bluvstein, Mikhail~D. Lukin, Alexander
  Lukin, Hengyun Zhou, and Sergio~H. Cantú.
\newblock Experimental demonstration of logical magic state distillation, 2024.

\bibitem{martinez2016compiling}
Esteban~A Martinez, Thomas Monz, Daniel Nigg, Philipp Schindler, and Rainer
  Blatt.
\newblock Compiling quantum algorithms for architectures with multi-qubit
  gates.
\newblock {\em New Journal of Physics}, 18(6):063029, jun 2016.

\bibitem{shapira2023fast}
Yotam Shapira, Lee Peleg, David Schwerdt, Jonathan Nemirovsky, Nitzan Akerman,
  Ady Stern, Amit~Ben Kish, and Roee Ozeri.
\newblock Fast design and scaling of multi-qubit gates in large-scale
  trapped-ion quantum computers, 2023.

\bibitem{schwerdt2024scalable}
David Schwerdt, Lee Peleg, Yotam Shapira, Nadav Priel, Yanay Florshaim, Avram
  Gross, Ayelet Zalic, Gadi Afek, Nitzan Akerman, Ady Stern, Amit~Ben Kish, and
  Roee Ozeri.
\newblock Scalable architecture for trapped-ion quantum computing using rf
  traps and dynamic optical potentials, 2024.

\bibitem{shapira2023programmable}
Yotam Shapira, Jovan Markov, Nitzan Akerman, Ady Stern, and Roee Ozeri.
\newblock Programmable quantum simulations on a trapped-ions quantum computer
  with a global drive, 2023.

\bibitem{shapira2020theory}
Yotam Shapira, Ravid Shaniv, Tom Manovitz, Nitzan Akerman, Lee Peleg, Lior
  Gazit, Roee Ozeri, and Ady Stern.
\newblock Theory of robust multiqubit nonadiabatic gates for trapped ions.
\newblock {\em Phys. Rev. A}, 101:032330, Mar 2020.

\bibitem{grzesiak2020efficient}
Nikodem Grzesiak, Reinhold Bl{\"u}mel, Kenneth Wright, Kristin~M. Beck, Neal~C.
  Pisenti, Ming Li, Vandiver Chaplin, Jason~M. Amini, Shantanu Debnath, Jwo-Sy
  Chen, and Yunseong Nam.
\newblock Efficient arbitrary simultaneously entangling gates on a trapped-ion
  quantum computer.
\newblock {\em Nature Communications}, 11(1):2963, Jun 2020.

\bibitem{lu2023realization}
Yao Lu, Wentao Chen, Shuaining Zhang, Kuan Zhang, Jialiang Zhang, Jing-Ning
  Zhang, and Kihwan Kim.
\newblock Realization of programmable ising models in a trapped-ion quantum
  simulator, 2023.

\bibitem{wu2023qubits}
Qiming Wu, Yue Shi, and Jiehang Zhang.
\newblock Qubits on programmable geometries with a trapped-ion quantum
  processor, 2023.

\bibitem{wang2022fast}
Kaizhao Wang, Jing-Fan Yu, Pengfei Wang, Chunyang Luan, Jing-Ning Zhang, and
  Kihwan Kim.
\newblock Fast multi-qubit global-entangling gates without individual
  addressing of trapped ions.
\newblock {\em Quantum Science and Technology}, 7(4):044005, aug 2022.

\bibitem{evered2023high}
Simon~J. Evered, Dolev Bluvstein, Marcin Kalinowski, Sepehr Ebadi, Tom
  Manovitz, Hengyun Zhou, Sophie~H. Li, Alexandra~A. Geim, Tout~T. Wang, Nishad
  Maskara, Harry Levine, Giulia Semeghini, Markus Greiner, Vladan
  Vuleti{\'{c}}, and Mikhail~D. Lukin.
\newblock High-fidelity parallel entangling gates on a neutral-atom quantum
  computer.
\newblock {\em Nature}, 622(7982):268--272, Oct 2023.

\bibitem{young2021asymmetric}
Jeremy~T. Young, Przemyslaw Bienias, Ron Belyansky, Adam~M. Kaufman, and
  Alexey~V. Gorshkov.
\newblock Asymmetric blockade and multiqubit gates via dipole-dipole
  interactions.
\newblock {\em Phys. Rev. Lett.}, 127:120501, Sep 2021.

\bibitem{cooper2024graph}
Eric~S. Cooper, Philipp Kunkel, Avikar Periwal, and Monika Schleier-Smith.
\newblock Graph states of atomic ensembles engineered by photon-mediated
  entanglement.
\newblock {\em Nature Physics}, 20(5):770--775, May 2024.

\bibitem{bravyi2022constant}
Sergey Bravyi, Dmitri Maslov, and Yunseong Nam.
\newblock Constant-cost implementations of clifford operations and
  multiply-controlled gates using global interactions.
\newblock {\em Phys. Rev. Lett.}, 129:230501, 2022.

\bibitem{bassler2023synthesis}
Pascal Ba{\ss{}}ler, Matthias Zipper, Christopher Cedzich, Markus Heinrich,
  Patrick~H. Huber, Michael Johanning, and Martin Kliesch.
\newblock Synthesis of and compilation with time-optimal multi-qubit gates.
\newblock {\em {Quantum}}, 7:984, 2023.

\bibitem{yu2013five}
Nengkun Yu, Runyao Duan, and Mingsheng Ying.
\newblock Five two-qubit gates are necessary for implementing the toffoli gate.
\newblock {\em Phys. Rev. A}, 88:010304, Jul 2013.

\bibitem{SM}
Supplemental material, that includes proofs and examples.

\bibitem{moll2018quantum}
Nikolaj Moll, Panagiotis Barkoutsos, Lev~S Bishop, Jerry~M Chow, Andrew Cross,
  Daniel~J Egger, Stefan Filipp, Andreas Fuhrer, Jay~M Gambetta, Marc Ganzhorn,
  Abhinav Kandala, Antonio Mezzacapo, Peter Müller, Walter Riess, Gian Salis,
  John Smolin, Ivano Tavernelli, and Kristan Temme.
\newblock Quantum optimization using variational algorithms on near-term
  quantum devices.
\newblock {\em Quantum Science and Technology}, 3(3):030503, jun 2018.

\bibitem{cross2019validating}
Andrew~W Cross, Lev~S Bishop, Sarah Sheldon, Paul~D Nation, and Jay~M Gambetta.
\newblock Validating quantum computers using randomized model circuits.
\newblock {\em Physical Review A}, 100(3):032328, 2019.

\bibitem{baldwin2022reexamining}
Charles~H. Baldwin, Karl Mayer, Natalie~C. Brown, Ciar{\'{a}}n Ryan-Anderson,
  and David Hayes.
\newblock Re-examining the quantum volume test: {I}deal distributions, compiler
  optimizations, confidence intervals, and scalable resource estimations.
\newblock {\em {Quantum}}, 6:707, May 2022.

\bibitem{miller2022improved}
Keith Miller, Charles Broomfield, Ann Cox, Joe Kinast, and Brandon Rodenburg.
\newblock An improved volumetric metric for quantum computers via more
  representative quantum circuit shapes, 2022.

\bibitem{jurcevic2021demonstration}
Petar Jurcevic, Ali Javadi-Abhari, Lev~S Bishop, Isaac Lauer, Daniela~F
  Bogorin, Markus Brink, Lauren Capelluto, Oktay G{\"u}nl{\"u}k, Toshinari
  Itoko, Naoki Kanazawa, et~al.
\newblock Demonstration of quantum volume 64 on a superconducting quantum
  computing system.
\newblock {\em Quantum Science and Technology}, 6(2):025020, 2021.

\bibitem{moses2023race}
S.~A. Moses, C.~H. Baldwin, M.~S. Allman, R.~Ancona, L.~Ascarrunz, C.~Barnes,
  J.~Bartolotta, B.~Bjork, P.~Blanchard, M.~Bohn, J.~G. Bohnet, N.~C. Brown,
  N.~Q. Burdick, W.~C. Burton, S.~L. Campbell, J.~P. Campora, C.~Carron,
  J.~Chambers, J.~W. Chan, Y.~H. Chen, A.~Chernoguzov, E.~Chertkov, J.~Colina,
  J.~P. Curtis, R.~Daniel, M.~DeCross, D.~Deen, C.~Delaney, J.~M. Dreiling,
  C.~T. Ertsgaard, J.~Esposito, B.~Estey, M.~Fabrikant, C.~Figgatt, C.~Foltz,
  M.~Foss-Feig, D.~Francois, J.~P. Gaebler, T.~M. Gatterman, C.~N. Gilbreth,
  J.~Giles, E.~Glynn, A.~Hall, A.~M. Hankin, A.~Hansen, D.~Hayes, B.~Higashi,
  I.~M. Hoffman, B.~Horning, J.~J. Hout, R.~Jacobs, J.~Johansen, L.~Jones,
  J.~Karcz, T.~Klein, P.~Lauria, P.~Lee, D.~Liefer, S.~T. Lu, D.~Lucchetti,
  C.~Lytle, A.~Malm, M.~Matheny, B.~Mathewson, K.~Mayer, D.~B. Miller,
  M.~Mills, B.~Neyenhuis, L.~Nugent, S.~Olson, J.~Parks, G.~N. Price, Z.~Price,
  M.~Pugh, A.~Ransford, A.~P. Reed, C.~Roman, M.~Rowe, C.~Ryan-Anderson,
  S.~Sanders, J.~Sedlacek, P.~Shevchuk, P.~Siegfried, T.~Skripka, B.~Spaun,
  R.~T. Sprenkle, R.~P. Stutz, M.~Swallows, R.~I. Tobey, A.~Tran, T.~Tran,
  E.~Vogt, C.~Volin, J.~Walker, A.~M. Zolot, and J.~M. Pino.
\newblock A race-track trapped-ion quantum processor.
\newblock {\em Phys. Rev. X}, 13:041052, Dec 2023.

\bibitem{quantinuum2024quantinuum}
Quantinuum.
\newblock Quantinuum hardware quantum volume data, 2024.
\newblock Commit : 76298b9.

\bibitem{aaronson2016complexity}
Scott Aaronson and Lijie Chen.
\newblock Complexity-theoretic foundations of quantum supremacy experiments,
  2016.

\bibitem{pino2021demonstration}
J.~M. Pino, J.~M. Dreiling, C.~Figgatt, J.~P. Gaebler, S.~A. Moses, M.~S.
  Allman, C.~H. Baldwin, M.~Foss-Feig, D.~Hayes, K.~Mayer, C.~Ryan-Anderson,
  and B.~Neyenhuis.
\newblock Demonstration of the trapped-ion quantum ccd computer architecture.
\newblock {\em Nature}, 592(7853):209--213, Apr 2021.

\bibitem{moses2023racetrack}
S.~A. Moses, C.~H. Baldwin, M.~S. Allman, R.~Ancona, L.~Ascarrunz, C.~Barnes,
  J.~Bartolotta, B.~Bjork, P.~Blanchard, M.~Bohn, J.~G. Bohnet, N.~C. Brown,
  N.~Q. Burdick, W.~C. Burton, S.~L. Campbell, J.~P. Campora, C.~Carron,
  J.~Chambers, J.~W. Chan, Y.~H. Chen, A.~Chernoguzov, E.~Chertkov, J.~Colina,
  J.~P. Curtis, R.~Daniel, M.~DeCross, D.~Deen, C.~Delaney, J.~M. Dreiling,
  C.~T. Ertsgaard, J.~Esposito, B.~Estey, M.~Fabrikant, C.~Figgatt, C.~Foltz,
  M.~Foss-Feig, D.~Francois, J.~P. Gaebler, T.~M. Gatterman, C.~N. Gilbreth,
  J.~Giles, E.~Glynn, A.~Hall, A.~M. Hankin, A.~Hansen, D.~Hayes, B.~Higashi,
  I.~M. Hoffman, B.~Horning, J.~J. Hout, R.~Jacobs, J.~Johansen, L.~Jones,
  J.~Karcz, T.~Klein, P.~Lauria, P.~Lee, D.~Liefer, S.~T. Lu, D.~Lucchetti,
  C.~Lytle, A.~Malm, M.~Matheny, B.~Mathewson, K.~Mayer, D.~B. Miller,
  M.~Mills, B.~Neyenhuis, L.~Nugent, S.~Olson, J.~Parks, G.~N. Price, Z.~Price,
  M.~Pugh, A.~Ransford, A.~P. Reed, C.~Roman, M.~Rowe, C.~Ryan-Anderson,
  S.~Sanders, J.~Sedlacek, P.~Shevchuk, P.~Siegfried, T.~Skripka, B.~Spaun,
  R.~T. Sprenkle, R.~P. Stutz, M.~Swallows, R.~I. Tobey, A.~Tran, T.~Tran,
  E.~Vogt, C.~Volin, J.~Walker, A.~M. Zolot, and J.~M. Pino.
\newblock A race-track trapped-ion quantum processor.
\newblock {\em Phys. Rev. X}, 13:041052, Dec 2023.

\bibitem{bluvstein2022quantum}
Dolev Bluvstein, Harry Levine, Giulia Semeghini, Tout~T. Wang, Sepehr Ebadi,
  Marcin Kalinowski, Alexander Keesling, Nishad Maskara, Hannes Pichler, Markus
  Greiner, Vladan Vuleti{\'{c}}, and Mikhail~D. Lukin.
\newblock A quantum processor based on coherent transport of entangled atom
  arrays.
\newblock {\em Nature}, 604(7906):451--456, Apr 2022.

\bibitem{khaneja2001time}
Navin Khaneja, Roger Brockett, and Steffen~J. Glaser.
\newblock Time optimal control in spin systems.
\newblock {\em Phys. Rev. A}, 63:032308, Feb 2001.

\bibitem{tucci2005introduction}
Robert~R. Tucci.
\newblock An introduction to cartan's kak decomposition for qc programmers,
  2005.

\bibitem{blaauboer2008analytical}
M~Blaauboer and RL~De~Visser.
\newblock An analytical decomposition protocol for optimal implementation of
  two-qubit entangling gates.
\newblock {\em Journal of Physics A: Mathematical and Theoretical},
  41(39):395307, 2008.

\bibitem{zhang2003geometric}
Jun Zhang, Jiri Vala, Shankar Sastry, and K.~Birgitta Whaley.
\newblock Geometric theory of nonlocal two-qubit operations.
\newblock {\em Phys. Rev. A}, 67:042313, Apr 2003.

\bibitem{ballance2016high}
C.~J. Ballance, T.~P. Harty, N.~M. Linke, M.~A. Sepiol, and D.~M. Lucas.
\newblock High-fidelity quantum logic gates using trapped-ion hyperfine qubits.
\newblock {\em Phys. Rev. Lett.}, 117:060504, Aug 2016.

\bibitem{smith2024single}
M.~C. Smith, A.~D. Leu, K.~Miyanishi, M.~F. Gely, and D.~M. Lucas.
\newblock Single-qubit gates with errors at the $10^{-7}$ level, 2024.

\bibitem{cuda-q}
{The CUDA-Q development team}.
\newblock Cuda-q, 2024.

\bibitem{crooks2024quantum}
Gavin~E Crooks.
\newblock Quantum gates.
\newblock page~26, 2024.

\end{thebibliography}
